\documentclass[apj]{emulateapj}
\usepackage{apjfonts}
\usepackage{amssymb}
\usepackage{graphicx}

\makeatletter

\usepackage{amsmath}

\def\araa{{\em ARA\&A}}
\def\ao{{\em Appl.\ Opt.\ }}
\def\aap{{\em A\&A}}

\def\apj{{\em ApJ}}
\def\apjl{{\em ApJL}}
\def\apjs{{\em ApJS}}

\def\mnras{{\em MNRAS}}
\def\nat{{\em Nature }}

\def\prd{{\em Phys.\ Rev.\ D }}

\makeatother

\begin{document}

\title{Spatially Extended 21 cm Signal from Strongly Clustered UV and X-Ray
Sources in the Early Universe}

\author{Kyungjin Ahn\altaffilmark{1}}
\author{Hao Xu\altaffilmark{2}}
\author{Michael L. Norman\altaffilmark{2}}
\author{Marcelo A. Alvarez\altaffilmark{3}}
\author{John H. Wise\altaffilmark{4}}

\altaffiltext{1}{Department of Earth Sciences, Chosun University,
    Gwangju, 501-759, Korea; kjahn@chosun.ac.kr}
\altaffiltext{2}{Center for Astrophysics and Space Sciences, University of
  California, San Diego, 9500 Gilman Drive, La Jolla, CA 92093}
\altaffiltext{3}{Canadian Institute for Theoretical Astrophysics, 60 St. George St.,
  Toronto, ON. M5S 3H8, Canada}
\altaffiltext{4}{Center for Relativistic Astrophysics, School of Physics,
  Georgia Institute of Technology, 837 State Street, Atlanta, GA 30332}

\keywords{cosmology:dark ages, reionization, first stars -- methods:
  numerical --  galaxy:high-redshift -- X-rays:galaxies -- radio lines:general}

\begin{abstract}
We present our prediction for the local 21 cm differential brightness
temperature ($\delta T_{b}$) from a set of strongly clustered sources
of Population III (Pop III) and II (Pop II) objects in the early Universe, by a numerical
simulation of their formation and radiative feedback. These objects
are located inside a highly biased environment, which is a rare, high-density
peak (``Rarepeak'') extending to $\sim7$ comoving Mpc. We study
the impact of ultraviolet (UV) and X-ray photons on the intergalactic
medium (IGM) and the resulting $\delta T_{b}$, when Pop III
stars are assumed to emit X-ray photons by forming X-ray binaries
very efficiently. We parameterize the rest-frame spectral energy
distribution (SED) of X-ray photons, which regulates X-ray photon-trapping,
IGM-heating, secondary Lyman-alpha pumping and the resulting morphology
of $\delta T_{b}$. A combination of emission ($\delta T_{b}>0$) and
absorption ($\delta T_{b}<0$) regions appears in varying amplitudes
and angular scales. The boost of the signal by the high-density environment
($\delta\sim0.64$) and on a relatively large scale combine
to make Rarepeak a discernible, spatially-extended ($\theta\sim10'$)
object for 21 cm observation at $13\lesssim z\lesssim17$, which is
found to be detectable as a single object by SKA with integration
time of $\sim1000$ hours. Power spectrum analysis by some of the SKA precursors
(LOFAR, MWA, PAPER) of such rare peaks is found difficult
due to the rarity of these peaks, and the contribution only
  by these rare peaks to the total power spectrum remains subdominant
  compared to that by all astrophysical sources.
\end{abstract}

\section{Introduction}
\label{sec:Introduction}

The first stars formed in the primordial chemical environment
of the early Universe, when all baryons were remnants of the big bang
nucleosynthesis. The metallicity was zero (except for a trace amount of
lithium and beryllium), and therefore the first stars are also identified
as zero-metallicity, or Pop III, stars. These stars were
born primarily inside minihalos, the first collapsed cosmological
halos with $T_{{\rm vir}}\lesssim10^{4}\,{\rm K}$. The formation,
evolution and death of these stars gradually enriched the environment
with metals, allowing the formation of Pop II stars when metallicity
reached about $10^{-4}$ of solar metallicity (see \citealt{Bromm2011}
and references therein). Both Pop II
and III stars emit ultraviolet photons capable of ionizing the surrounding
hydrogen and helium atoms, and thus  early formation epoch also
marks the cosmic dawn (CD) and the beginning of the epoch of reionization
(EoR).
The evolution of this high-redshift astrophysics is marked
conveniently by three prominent epochs: (1) the Ly$\alpha$ -pumping
epoch, when the IGM is strongly coupled to the gas kinetic temperature ($T_{k}$)
through the Wouthysen-Field effect with high Ly$\alpha$
intensity, (2) the X-ray heating epoch, when the IGM is heated to beyond $T_{\rm CMB}$,
the temperature of the cosmic microwave background (CMB) by the
X-ray heating and (3) the main EoR, when H II bubbles in cosmological
scales form in a patchy way. 
It is generally believed that the CD commences with the the Ly$\alpha$-pumping epoch, followed by
the X-ray heating epoch, and finally occurs the EoR, whose sequence is rather robust unless 1-2
order-of-magnitude changes in the fiducial astrophysical parameters
are allowed
\citep{2006MNRAS.371..867F,McQuinn2012,Mesinger2014}. While this might
be true, the recent development in the theory of the first-star
formation and the subsequent X-ray-source formation brings a new
possibility that the X-ray heating epoch may have started 
earlier than previously thought, as we describe below.

Early theoretical work on the formation of the first stars, mostly through
high resolution simulations, found that Pop III stars with mass
$M_{{\rm III,*}}\gtrsim100\, M_{\odot}$ are born in isolation inside minihalos,
and thus the ``one massive Pop III star per minihalo'' paradigm was
established (\citealt{2002Sci...295...93A}; \citealt{2002ApJ...564...23B};
\citealt{Yoshida2006}). Later, however, several higher-resolution
simulations began to observe the formation of binary protostar systems with
a smaller mass range, $M_{{\rm III},*}\simeq[10-40]\, M_{\odot}$ (\citealt{Turk2009};
\citealt{Stacy2010}). While the universality of
the latter finding is in doubt
(\citealt{Greif2012}; \citealt{Stacy2013}; 
\citealt{Susa2013}; \citealt{Hirano2014};  see also
  \citealt{Becerra2015} for the 
formation of Pop III stars inside more massive halos), 
it certainly introduces very important 
subtleties to the old paradigm. One important aspect is that X-ray binary
systems may remain after some of these stars die, and can emit X-rays
very efficiently (we will quantify their X-ray emissivity in
Section~\ref{sec:21cm_rarepeak}) through gas accretion rate comparable
to the Eddington 
limit (\citealt{Mirabel2011}). 
This then could make the X-ray heating epoch occur earlier than
previously thought, or even allow a model where the reionization is
dominated by X-ray photons instead of UV photons.
Reionization dominated by X-ray photons,
due to their long mean free path, will occur much more smoothly in
space than reionization by UV photons (e.g. \citealt{Mesinger2013};
\citealt{Haiman2011} and references therein).
In addition, X-rays can heat the IGM (see e.g. recent
observational signature reported by \citealt{Parsons2014}), which
impacts the dynamics of the IGM (e.g. 
\citealt{Tanaka2012}; \citealt{Jeon2014}) and $\delta T_{b}$ of IGM
(e.g. \citealt{Mesinger2013}; \citealt{Fialkov2014};
\citealt{Jeon2014}). High-redshift X-ray binaries seem to 
dominate the X-ray background over the active galactic nuclei at the
late stage of EoR ($z\sim 6-8$), if their emissivity and SED
is calibrated by the observed, low-$z$ ($0\le
z\lesssim 4$) X-ray binaries \citep{Fragos2013}.

It is important to study the observational aspect of this new
scenario of the formation and evolution of the Pop III
objects during the CD. Because of short lifetime of these objects (e.g. Pop III
stars with mass $\gtrsim 100\,M_\odot$ live for less than a few million
years), direct observation should aim a very high redshift range.
Among many, one of the most sought-after
probes of such high redshift objects is the observation of the redshifted
21 cm line from neutral hydrogen, because it can probe the IGM structure and the impact from the
early radiation sources simultaneously. This is the main science goal of the next-generation radio
telescopes (e.g. LOFAR - LOw Frequency ARray, MWA - Murchison
Widefield Array, PAPER - Precision Array for Probing the Epoch of
Reionization, HERA - The Hydrogen Epoch of Reionization Array, SKA -
Square Kilometre Array, etc.). 
The 21 cm observation will be driven mainly in two ways: the power spectrum
analysis and the real-space tomography (3D imaging). The power
spectrum analysis has the merit of achieving relatively high
sensitivity by stacking many wavemodes at an equal
radius in the Fourier space (\citealt{Mellema2013} and references therein), and thus is
possible with SKA precursors (LOFAR, MWA, PAPER, HERA). Using the
anisotropy due to the redshift-space distortion, the power spectrum
analysis can even allow separation of the cosmological
information from the astrophysical information (BL; \citealt{Mao2012}:
MSMIKA hereafter) during the 
early epoch of cosmic reionization. Nevertheless, the real-space
tomography should be carried out eventually to obtain more information
for the generically non-Gaussian field, which will become practical
with SKA, the highest-sensitivity radio apparatus.

Predictions have been made on the possible tomography of individual high-redshift
objects during the CD, which is also the focus of this paper.  When UV sources are embedded in
the IGM colder than the CMB, a ``Ly$\alpha$ sphere'' forms around them
and the 21-cm absorption trough forms. When these UV sources are
accompanied by X-ray sources, the central regions is heated to be
observed in
emission but is still surrounded by the absorption trough against the
CMB continuum (e.g. \citealt{2000ApJ...528..597T}; \citealt{Cen2006};
\citealt{2006ApJ...648L...1C}; \citealt{Chen2008};
\citealt{Alvarez2010}). Because this feature is not
likely to form during the main EOR phase when X-ray heating is
efficient everywhere and the patchy H II regions dominate the signal,
the Ly$\alpha$ sphere can be a smoking gun for the very early
astrophysical objects. Tomography of the signature of a single Pop III star
(e.g. \citealt{Cen2006}; \citealt{Chen2008}) is
practically impossible at this stage even with SKA, because its zone
of influence on the IGM is too small to be observed with reasonable
sensitivity. Observing a quasar system with a supermassive black hole
\citep{2000ApJ...528..597T,Alvarez2010} seems more promising,
because the zone of influence is much more extended than that of a
single star and thus guarantees much better detectability. Nevertheless,
the number density of high-redshift quasar systems are calculated
under an ad-hoc assumption \citep{Alvarez2010}, and thus the required volume of
observation (field-of-view $\times$ redshift-range) is uncertain.

In this paper, instead of a single star, galaxy, or a quasar system,
we examine a system of highly clustered galaxies hosting Pop III and II stars together
with the X-ray binaries, which have many observational merits as follows.
Rare, high-density peaks in the Universe are a good
site for Pop III and II stars to form and get clustered, and the
number density of such peaks can be easily calculated under a fixed
density-filtering scale. 
Because they form much earlier than
more average peaks under a given filtering (or mass) scale, these high-density
peaks may stand out as almost-isolated objects until other
smaller-scale objects start to become abundant. Moreover, strong
clustering of radiation sources may extend their zone of influence to
the extent that is observable by the 21 cm tomography. How the impact
of X-ray binaries associated with the death of binary Pop III stars
will be seen in 21 cm is also of prime interest, as it involves a
new development in the theory of first star formation and may
affect how cosmic reionization progressed.
Observation of such a signature will allow constraining a few physical
parameters of high-redshift radiation sources, such as their spectral
energy distribution, UV and X-ray emissivity, and their clustering
scales. For example, \citet{Fialkov2014} studied the impact of the SED
of high-redshift X-ray binaries, and concluded that a very hard SED case would result
in very uniform heating of the IGM, such that the resulting 21-cm
fluctuation during the X-ray heating epoch may be much smaller than
previously estimated. 
High-redshift X-ray binaries surely increase model uncertainties
  and can affect the high-redshift 21-cm observations \citep{Fialkov2014a}.

We study the observational signature of highly clustered Population
III and II stars and Pop III X-ray binaries inside ``Rarepeak'',
a rare and high-density environment, within which the formation and
evolution of both Pop III and II stars and their UV radiative
feedback on the IGM has been simulated up to $z=15$ inside a $40\,{\rm Mpc}$
box (\citealt{Xu2013}, XWN hereafter), with a particular focus on the 21-cm differential
brightness temperature with respect to the cosmic microwave background. Radiation from Pop III
X-ray binaries was included in this box in postprocessing, in order to determine
the temperature and ionization states of IGM (\citealt{Xu2014}, XAWNO
hereafter), which we use to calculate $\delta T_{b}$. While the 21-cm
signature of individual sources has been studied for individual or
composite first stars, galaxies and black hole systems (\citealt{2000ApJ...528..597T};
\citealt{Cen2006}; \citealt{Chuzhoy2006}; \citealt{Chen2008}; \citealt{Alvarez2010}),
Rarepeak is unique in that (1) it is much more realistic than the
idealized, spherically-symmetric geometries considered in previous work, (2)
it has a very large volume, $\sim150\,{\rm Mpc}^{3}$ comoving, and
mass, $M\sim8.3\times10^{12}\, M_{\odot}$, as a single clustered
object at $z=15$ which would seed a small proto galaxy cluster later
(for example, at $z=6$, two $\sim3\times10^{10}\, M_{\odot}$ halos
appear inside Rarepeak in addition to many smaller galaxies) and (3)
has a relatively large mean overdensity, $\left\langle \delta\right\rangle \sim0.64$
at $z\sim15$, which amplifies the 21-cm signal. We investigate
whether the last two factors will combine to open a new observational
window for detecting high-redshift objects. 

As shown in \citet{2000ApJ...528..597T}, \citet{Cen2006}, \citet{Chuzhoy2006},
and \citet{Chen2008}, the existence of a Ly$\alpha$ sphere with
a strong absorption trough of $\delta T_{b}\sim-100\,{\rm mK}$ around
these sources seems ubiquitous. However, X-ray binary systems may
be very efficient in heating the IGM before the absorption trough occurs,
which we also investigate here. This paper is organized as follows. In
Section \ref{sec:simulation}, we briefly describe the numerical
simulation of Rarepeak and how we calculate the inhomogeneous
Ly$\alpha$ and X-ray background, which determine $\delta
T_{b}$. We also describe the  parametrization of 
the X-ray SED. In Section \ref{sec:21cm_rarepeak}, we describe the
characteristics of the calculated $\delta T_{b}$ field.
In Section \ref{sec:Detectability}, we present our forecasts for 21-cm
observations of Rarepeak, by tomography (Section \ref{sub:Tomography})
and power spectrum 
analysis (Section \ref{sub:Power-spectrum-analysis}). We conclude this
work with a summary and 
a discussion on observational prospects and some concerns on
high-redshift 21-cm cosmology with regard to our result in Section
\ref{discussion}. Appendix is added to describe the detailed scheme
we developed and used for calculating $\delta T_{b}$ inside the
observing data cube for the 
most generic cases with finite optical depth and peculiar velocity.

\section{Simulation of a rare density peak and radiation transfer}
\label{sec:simulation}

\subsection{Radiation Hydrodynamics Simulation of Rarepeak and
  Formation of X-ray Binaries}
\label{sec:sim_rarepeak}
Rarepeak is a local density maximum whose mean density is significantly
larger than the average density, and thus biased formation of Population
III and II stars occurs at all times. XWN performed a simulation
of the formation and UV radiation feedback of Pop III and II stars
in Rarepeak from $z=99$ to $z=15$, using the adaptive mesh refinement
(AMR) code Enzo (\citealt{Bryan2014}) and ray-tracing method for UV
transfer  Moray (\citealt{Wise2011}). A cubic, comoving, periodic box of volume $(40\,{\rm Mpc})^{3}$
was used, with a $512^3$ root grid
resolution and three levels of static nested grids centered on this
high density region (see XWN for details). The first run was a pure
N-body simulation with $512^3$ particles, which ran from $z=99$ to
$z=6$. Then, a Lagrangian volume 
(cuboid) containing two $\sim 3\times 10^{10} \,M_\odot$ halos at
$z=6$ was selected, and the simulation restarted now with three more
static nested grids to have an effective resolution of $4096^3$ and
an effective dark matter mass resolution of $2.9\times 10^4
\,M_\odot$ inside the highest nested grid covering a comoving volume
of $5.2\times 7.0\times 8.3\,{\rm Mpc}^3$. We call this region
``Rarepeak''. 
Inside Rarepeak, identifying halos with $\sim 50$ or
more dark matter particles, the minimum halo mass resolved is $\sim
10^{6} \,M_\odot$. This halo mass resolution seems good enough to
cover the whole range of star-forming halos: a finer-resolution
simulation with the minimum halo mass of $\sim 2 \times
10^{5}\,M_\odot$ still finds that stars are forming only inside halos
of mass greater than $\sim 3\times 10^{6}\,M_\odot$ \citep{Wise2012}.
Depending upon the
refinement criteria \citep{Wise2012}, a maximum refinement of level
$l=12$ was allowed, resulting in a maximum resolution of 19 comoving
pc. This Lagrangian volume has, at $z=15$, a comoving volume of $\sim
138\,{\rm Mpc}^3$ which is a $3.45\sigma$ density peak under the
corresponding filtering scale (see also
Section~\ref{sub:Power-spectrum-analysis}). At this time, this
region contains more than 10000 Pop III stars and remnants
distributed over 3000 halos, most of which are more massive than
$10^7 \,M_\odot$.

Together with the simulation of the structure formation, the transfer
of H- and He-ionizing photons from all the Pop III and Pop II stars
and calculating the ionization fractions with a rate solver are all carried out
simultaneously. Hydrodynamics is also carried out self-consistently to
follow the evolution of density, temperature and velocity fields.
In XWN and XAWNO, a star particle is formed when
the star formation criterion is met (see XWN, XAWNO and \citealt{Wise2012} for details).
If a grid cell containing the star particle has metallicity
[Z/H]$<-4$, the particle becomes a Pop III star (or a Pop III binary
system) whose mass is a
randomly chosen sample from an initial mass function (IMF) given by
\begin{equation}
f(\log M)dM=M^{-1.3}\exp\left[-\left(\frac{M_{\rm char}}{M}\right)^{1.6}\right]dM,
\label{eq:PopIII_IMF}
\end{equation}
and the star particle becomes a Pop II stellar cluster otherwise. Here the
characteristic mass for the Pop III IMF, $M_{\rm char}$, is taken as
$40\,M_\odot$.

Based on this simulation, we calculate the impact of X-ray binaries as follows.
First, we assume that 50\% of the Pop III star particles form
X-ray binary systems. Then, the initial black hole is assigned a
mass of $40\,M_\odot$ or $10\,M_\odot$, if the mass of the
star particle satisfies $M_{*}/M_\odot>40$ or $10<M_{*}/M_\odot<40$, respectively.
We further assume that the black hole then accrets matter of the
companion star at the Eddington limit during the lifetime of the star $\tau_*$,
such that the luminosity becomes
\begin{equation}
L=L_{\rm Edd}=1.3\times 10^{38}\,\left(M_{\rm BH}/M_{\odot}\right){\rm erg/s},
\label{eq:LEdd}
\end{equation}
with radiation efficiency $\epsilon\equiv L/\dot{M}c^2$ and the black
hole mass grows from the initial value, $M_{{\rm BH},0}$, by
\begin{equation}
M_{\rm BH}=M_{{\rm BH},0} {\rm e}^{t/t_{\rm Edd}},
\label{eq:MBH}
\end{equation}
where the Eddington time 
$t_{\rm Edd}=M_{\rm BH}/\dot{M}_{\rm Edd}\sim 440\epsilon \,{\rm Myr}$. 
The total accreted mass becomes 
${\rm min}(M_{*},\,M_{{\rm BH},0}{\rm e}^{\tau/t_{\rm Edd}})-M_{{\rm BH},0}$.
The under-resolved region outside Rarepeak, which in fact takes almost
all the simulation volume, is populated by UV and X-ray sources as
follows. Inside Rarepeak, we project the UV (X-ray) luminosity and baryon
density to the root grid of $512^3$ cells. Among all the cells, we
sample only those containing UV (X-ray) sources and obtain the average
UV (X-ray) luminosity $\bar{L}$  and the baryon density
$\bar{\rho}$. Then, all the cells with density higher than $\bar{\rho}$
are assigned the UV (X-ray) luminosity of $\bar{L}$. The net
luminosity of the simulation box is found to be dominated by these new
sources over the Rarepeak by a factor of $\sim 3 - 9$ depending on the
redshift. Updating the net UV and X-ray luminosities this way are very
important in estimating the Ly$\alpha$ background intensity and X-ray
heating rate + secondary Ly$\alpha$ pumping, respectively (Section~\ref{sec:RT_X_Lya}).

We assume two types of rest-frame SEDs
for X-rays. The first type is a simple monochromatic SED, given by
luminosity $L_{\nu_{s}}=L_{{\rm X},\,0}\delta^{D}(\nu_{s}-\nu_{0})$
where $L_{{\rm X},\,0}$ (${\rm erg\, s^{-1}}$) is the bolometric
X-ray luminosity and is multiplied by the Dirac delta function centered
at frequency $\nu_{0}$. We use a constant $L_{{\rm X},\,0}$ such
that the number of X-ray photons is inversely proportional to $\nu_{0}$.
The second type is a composite power-law SED
\begin{equation}
L_{\nu_{s}}\propto\begin{cases}
\nu_{s} & \,\,\,\,{\rm if}\,\, h\nu_{s}<400\,{\rm eV},\\
\nu_{s}^{-1} & \,\,\,\,{\rm if}\,\,400\,{\rm eV}<h\nu_{s}<10\,{\rm keV},
\end{cases}\label{eq:power-law-SED}
\end{equation}
where again the bolometric luminosity $\int d\nu_{s}L_{\nu_{s}}$
is set to equal the constant value $L_{{\rm X},\,0}$. For the first
type, we parametrize $\nu_{0}$ to observe the impact of the X-ray
energy: $\epsilon_{{\rm X},0}\equiv h\nu_{0}=$\{0.3, 0.5, 0.77, 1,
3\} keV. For the second type, we use five bands of above
frequencies $\{\nu_{0}\}$ with luminosities weighted by \{0.134,
0.125, 0.089, 0.219, 0.433\}, respectively. The photo-heating and
photo-ionization by X-ray photons are calculated by correctly implementing
the build-up of X-ray background.

\subsection{Transfer of X-ray and Ly$\alpha$ radiation}
\label{sec:RT_X_Lya}
We calculate all quantities locally, by calculating physical quantities
within the entire simulated volume on a $256{}^{3}$-cell uniform grid. We
explicitly calculate the two crucial, locally-varying, radiation fields:
the proper Ly$\alpha$ number intensity $N_{\alpha}$ (${\rm cm^{-2}\, s^{-1}\, Hz^{-1}\, sr^{-1}}$)
and the proper X-ray intensity $J_{{\rm X},\nu_{{\rm obs}}}$ (${\rm erg\, cm^{-2}\, s^{-1}\, Hz^{-1}\, sr^{-1}}$)
as follows (similar to the calculation of Lyman-Werner background
by \citealt{Ahn2009}). First, Ly$\alpha$ photons can be generated
both by redshifted UV continuum below the Lyman limit frequency and by
collisional excitation of HI atoms heated by X-ray photons: let us
denote the number intensity of the former by $N_{\alpha,{\rm UV}}$
and of the latter by $N_{\alpha,{\rm X}}$. A point source with the
rest-frame photon number luminosity $N_{\nu_{s}}\,({\rm s^{-1}\, Hz^{-1}})$
at the source frequency $\nu_{s}$ (below the Lyman limit), the comoving
coordinate position ${\bf x'}$ and the source redshift $z_{s}$ will
generate, at a comoving coordinate position ${\bf x}$ and an observing
redshift $z_{{\rm obs}}$,
\begin{eqnarray}
&&N_{\alpha,{\rm UV}}({\bf x},\, z_{{\rm obs}};\,{\bf x'},\,
  z_{s})=\sum_{n=2}^{n_{{\rm max}}}\Theta(\nu_{n+1}-\nu_{s}(n))
  \nonumber \\
&&\times f_{{\rm recycle}}(n)\frac{N_{\nu_{s}(n)}({\bf x'},\, z_{s})}{(4\pi)^{2}D_{L}^{2}(z_{{\rm obs}},\, z_{s})}\left(\frac{1+z_{s}}{1+z_{{\rm obs}}}\right)^{2},\label{eq:Nalpha}
\end{eqnarray}
where  $n$ is the principal quantum number and the luminosity distance $D_{L}$ is given in terms of the comoving
distance $r_{{\rm os}}\equiv\left|{\bf x'}-{\bf x}\right|=2cH_{0}^{-1}\Omega_{m}^{-0.5}\left[(1+z_{{\rm obs}})^{-0.5}-(1+z_{s})^{-0.5}\right]$
as
\begin{equation}
D_{L}(z_{{\rm obs}},\, z_{s})=\left(\frac{r_{{\rm os}}}{1+z_{{\rm obs}}}\right)\left(\frac{1+z_{s}}{1+z_{{\rm obs}}}\right).\label{eq:DL}
\end{equation}
In the above, $f_{{\rm recycle}}(n)$ is the recycling rate of a source
photon with $\nu_{s}=\nu_{s}(n)\equiv\nu_{n}({1+z_{s}})/({1+z_{{\rm obs}}})$,
when it is redshifted to the nearest Lyman resonance frequency $\nu_{n}\equiv\nu_{\alpha}({4}/{3})(1-{1}/{n^{2}})=2.47\times10^{15}{\rm Hz}\,({4}/{3})(1-{1}/{n^{2}})$
at $\{{\bf x},\, z_{{\rm obs}}\}$ (we use $f_{{\rm recycle}}(n)$
calculated by \citealt{Pritchard2006}), being recycled to a Ly$\alpha$
photon by subsequent cascades and multiple scatterings. 
This is then
multiplied by the Heaviside function $\Theta$ ($\Theta(x)=1$ if
$x\ge0$; $\Theta(x)=0$ otherwise), indicating the redshift of $\nu_{s}(n)$
to $\nu_{n}$ at $z$ without being absorbed by the higher
resonance frequency $\nu_{n+1}$. This ``dark screen'' effect generates
the well-known step-wise profile of $N_{\alpha,{\rm UV}}(r_{{\rm os}})$
decreasing more rapidly than $1/D_{L}^{2}$. We truncate the summation
at $n_{{\rm max}}=23$, which is a rough estimation of the impact
of local ionization by a source (e.g. \citealt{2005ApJ...626....1B};
\citealt{Pritchard2006}). We then calculate the net intensity, $N_{\alpha,{\rm UV}}({\bf x},\, z_{{\rm obs}})=\sum_{{\bf x'}}N_{\alpha,{\rm UV}}({\bf x},\, z_{{\rm obs}};\,{\bf x'},\, z_{s})$,
by summing contributions from all the sources inside and outside the
box: we attach the simulated boxes periodically for the latter, and
we consider the time dilation such that boxes use progressively older
snapshots as look-back time increases. The calculation is accelerated
by using the fast-Fourier-transform scheme for discretized look-back
time contributions. We find that the out-of-box contribution becomes nearly uniform while the inhomogeneity of the radiation field is dominated by the inside-the-box
contribution. Of course, modes of fluctuation with wavelengths
  larger than the simulaton box is not this periodic, and realizing
  these modes in our simulation is impossible. Nevertheless, 
  the fluctuation of long mean-free-path radiation fields is usually
  dominated by nearby sources (e.g. Lyman-Werner fluctuation compiled
  by \citealt{Ahn2012}), and thus we simply use the scheme described above.

Second, we calculate the X-ray intensity field $J_{{\rm X},\nu_{{\rm obs}}}({\bf x},\, z_{{\rm obs}})$
in the same manner for time dilation and periodicity as in obtaining
$N_{\alpha,{\rm UV}}({\bf x},\, z_{{\rm obs}})$, by summing over the point-source
contribution
\begin{eqnarray}
J_{{\rm X},\nu_{{\rm obs}}}({\bf x},\, z_{{\rm obs}};\,{\bf x'},\,
z_{s})&=&\frac{L_{\nu_{s}}({\bf x'},\,
  z_{s})}{(4\pi)^{2}D_{L}^{2}(z_{{\rm obs}},\, z_{s})} \nonumber \\
& &\times \left(\frac{1+z_{s}}{1+z_{{\rm obs}}}\right)e^{-\tau},\label{eq:JX}
\end{eqnarray}
where $L_{\nu_{s}}({\bf x'},\, z_{s})$ is the source luminosity (${\rm erg\, s^{-1}\, Hz^{-1}}$)
in the X-ray band, and an appropriate calculation of the X-ray optical
depth $\tau$ is made (XAWNO). Finally, $N_{\alpha,{\rm X}}({\bf x},\, z_{{\rm obs}})$
is given by
\begin{eqnarray}
N_{\alpha,{\rm X}}({\bf x},\, z_{{\rm
    obs}})&=&\frac{c\eta_{\alpha}(x)}{H(z_{{\rm obs}})h\nu_{\alpha}^{2}}
\sum_{i} n_{i}\int_{\nu_{i}}^{\infty}d\nu_{{\rm obs}}\, h(\nu_{{\rm
    obs}}-\nu_{i}) \nonumber \\
& &\times \left(J_{{\rm X},\nu_{{\rm obs}}}({\bf x},\, z_{{\rm obs}})/h\nu_{{\rm obs}}\right)\sigma_{i}(\nu_{{\rm obs}}),\label{eq:2ndLya}
\end{eqnarray}
where $\eta_{\alpha}(x)=0.4766\left(1-x^{0.2735}\right)^{1.5221}$,
given only by local ionized fraction $x$, is the fraction of the
primary electron's energy going into the secondary excitation of HI
(\citealt{Shull1985}), $n_{i}$, $\nu_{i}$ and $\sigma_{i}$ are
the number density, the ionization energy and the absorption cross
section of species $i$ (=H I, He I, He II), respectively, $H(z_{{\rm obs}})$
is the Hubble constant at $z=z_{{\rm obs}}$, $h$ is the Planck constant,
and $\nu_{\alpha}$ is the HI Ly$\alpha$ frequency (we neglect He
II contribution though, due to low ionization states at $z\ge15$).
Equation (\ref{eq:2ndLya}) is the result of the fact that the originally
line-centered Ly$\alpha$ photons created by the secondary excitation
at a rate $\propto J_{{\rm X},\nu_{{\rm obs}}}$ is balanced by the
photons being redshifted out of the line center at a rate $H\nu_{\alpha}$
(\citealt{Chen2008}). We use the total intensity, $N_{\alpha}({\bf x},\, z_{{\rm obs}})=N_{\alpha,{\rm UV}}({\bf x},\, z_{{\rm obs}})+N_{\alpha,{\rm X}}({\bf x},\, z_{{\rm obs}})$,
to calculate Ly$\alpha$ pumping of the 21 cm lines. With the background
X-ray intensity $J_{{\rm X},\nu_{{\rm obs}}}({\bf x},\, z_{{\rm obs}})$,
which includes all the sources inside and outside the box by properly
treating the redshifting of photons and time dilation, we calculate
and ``update'' the kinetic temperature $T_{k}$ and the electron
fraction $x$ to include secondary ionizations (XAWNO).

\section{21-cm differential brightness temperature of Rarepeak region}
\label{sec:21cm_rarepeak}

$\delta T_{b}$ is determined by the spin temperature $T_{s}$, coupled
to the CMB temperature $T_{{\rm CMB}}$ and the kinetic temperature
of baryons $T_{k}$ through
\begin{equation}
T_{s}^{-1}=\frac{T_{{\rm CMB}}^{-1}+(x_{\alpha}+x_{c})T_{k}^{-1}}{1+x_{\alpha}+x_{c}},\label{eq:Tspin}
\end{equation}
where $x_{\alpha}=({16\pi^{2}T_{*}e^{2}f_{12}})/({27A_{10}T_{{\rm
      CMB}}m_{e}c})S_{\alpha} N_{\alpha}=S_{\alpha} N_{\alpha}/({1.165\times10^{-10}[(1+z)/20]}
  {\rm cm^{-2}\, s^{-1}\, Hz^{-1}\, sr^{-1}})$
is the Ly$\alpha$ coupling coefficient with a correction factor of
the order of unity $S_{\alpha}$  and the local Ly$\alpha$ photon
number intensity $N_{\alpha}$, and $x_{c}=({4\kappa_{1-0}(T_{k})n_{{\rm H}}T_{*}})/({3A_{10}T_{{\rm CMB}}})$
is the collisional coupling coefficient. 
Here, $T_{*}\equiv
hc/k\lambda_{21\rm cm}=0.0628\,{\rm K}$,
$A_{10}=2.85\times10^{-15}\,{\rm s^{-1}}$ is the spontaneous emission
coefficient, $f_{12}=0.4162$ is the oscillator strength for the
Ly$\alpha$ transition, and we use the tabulated
values for $\kappa_{1-0}(T_{k})$ by \citet{2005ApJ...622.1356Z}. In
this work we adopt the form suitable 
for comoving gas without peculiar velocity, $S_{\alpha}=\exp\left[-0.37(1+z)^{1/2}T_{{\rm K}}^{-2/3}\right]\left(1+{0.4}/{T_{{\rm K}}}\right)^{-1}$,
calculated by \citealt{Chuzhoy2006}, where $T_{{\rm K}}$ is the gas
temperature in Kelvin; for similar results see \citet{2004ApJ...602....1C}
and \citet{2006MNRAS.367..259H}.

We
accurately calculate $\delta T_{b}$ considering both the finite
optical depth and the peculiar velocity of gas elements (Appendix). 
In the highly nonlinear regime reached by
e.g. Rarepeak, some gas elements can be
optically thick and also move at high peculiar velocities. 
Large optical depth, where it occurs, invalidates the usual, optically thin
approximation, and 
high peculiar velocities may lead to significant shifts of the signal in
the observing frequency space. In addition, real-space cells may
overlap in the observing frequency space due to the spatial variance of the
peculiar velocity. Therefore, we need a new scheme different from the
typical optically thin approximation, which is described in Appendix.

\begin{figure*}
\begin{center}

\includegraphics[width=0.9\textwidth]{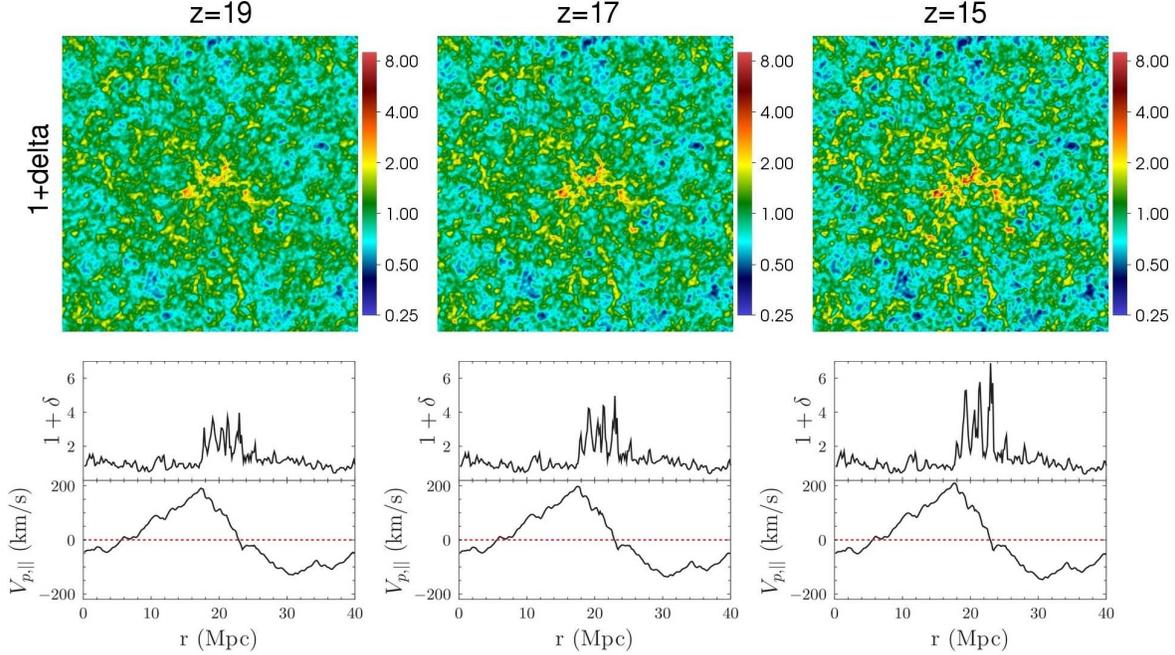}
\end{center}

\caption{(1st row) Distribution of hydrogen density ($1+\delta_{{\rm HI}}$)
at $z=$19, 17 and 15 (left to right), on a 2D slice (0.156 Mpc thick)
of the simulation box of $40$ 
comoving Mpc. All 2D maps in this paper are plotted in the full box
scale, and are centered around this slice when
frequency-averaging is taken.
Because the simulation ends at $z=15$, for later epochs we use
the density field frozen at $z=15$. 
(2nd row) Distribution of hydrogen density ($1+\delta_{{\rm HI}}$)
along one line of sight (z-axis or the frequency coordinate), which
pierces the center of each 2D
slice (xy-plane or the 2D sky coordinates) above
perpendicularly. Also plotted are the peculiar velocities
($v_{\parallel}$) projected 
along the line of sight. Increasing r corresponds to 
the far side of the box from the observer.
\label{fig:densvel}}
\end{figure*}

\begin{figure*}
\begin{center}
\includegraphics[width=0.90\textwidth]{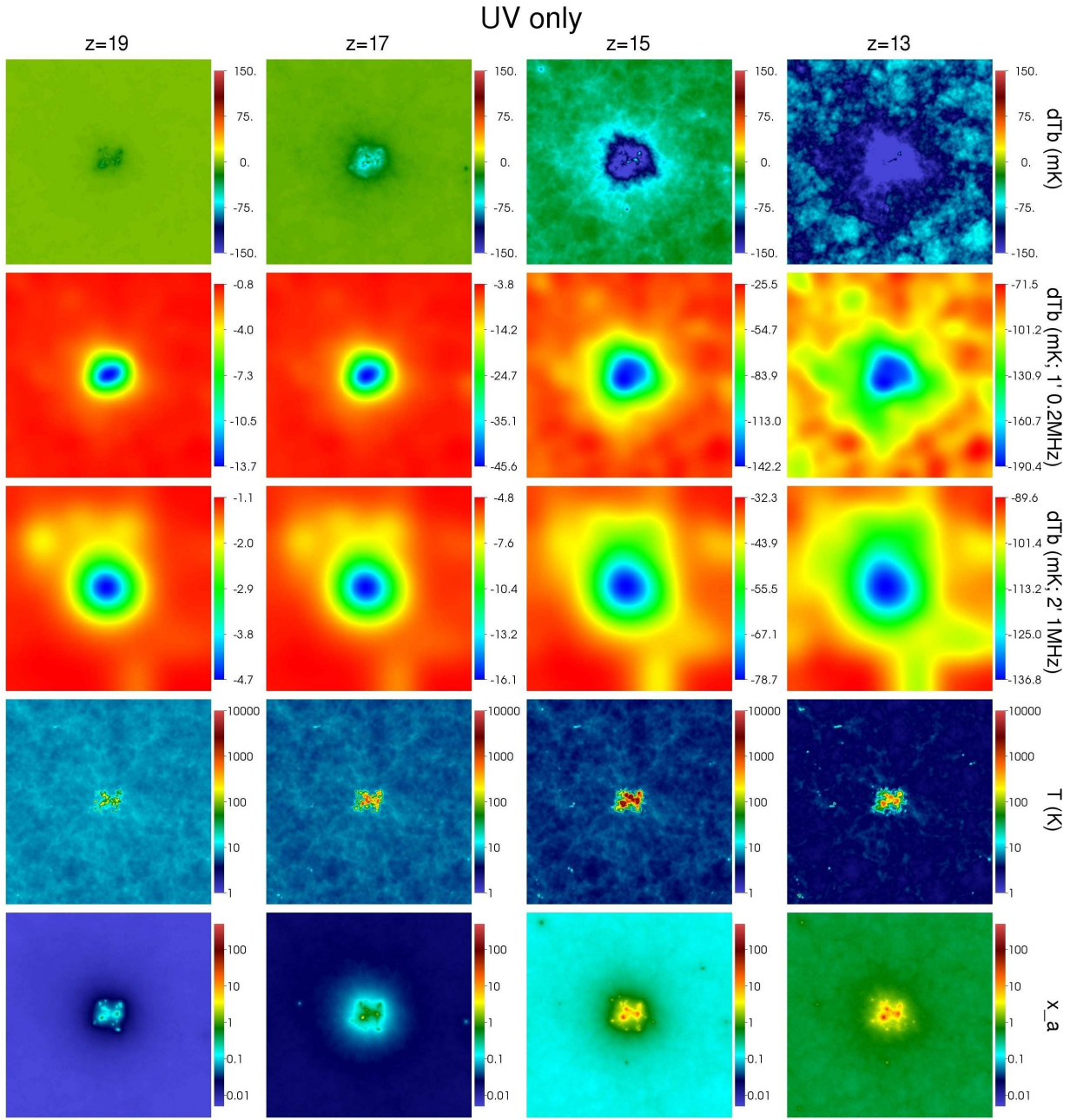}
\end{center}
 
\vspace{-0.2cm}
\caption{2D maps of 21 cm differential brightness temperature $\delta T_{b}$
(mK), gas temperature $T$ (K), and the Ly$\alpha$ coupling coefficient
$x_\alpha$, when no X-ray sources are present. From left to right,
each column corresponds to $z=$19, 17, 15, and 13,
respectively. From top to bottom, the 1st, 4th and 5th rows
represent $\delta T_{b}$, $T$ and $x_\alpha$, all of which are filtered only
along the frequency axis with $\Delta\nu= 0.2\,{\rm MHz}$. The 2nd and the
3rd rows represent $\delta T_{b}$ filtered with 
$\{\Theta,\,\Delta\nu\}=\{1',\,0.2\,{\rm MHz}\}$
and $\{\Theta,\,\Delta\nu\}=\{2',\,1\,{\rm MHz}\}$, respectively.
We apply carefully tuned, universal color schemes to the fields of
  $\delta T_{b}$, $T$ and $x_\alpha$ filtered only along the frequency
  axis (1st, 4th and 5th rows) throughout
  Figures~\ref{fig:UVonly}-\ref{fig:3keV}, in order to 
  demonstrate the model dependency clearly. Unfortunately, due to a
  strong model dependency, $\delta T_{b}$ in some
  models are shown in saturated color when the actual $\delta T_{b}$ is
  smaller than the minimum value in the color scheme, $\delta
   T_{b}=-150\,{\rm mK}$. For example, the actual
  minimum of $\delta T_{b}$ (1st row) at $z=13$ (4th column) is $-275.4\,
  {\rm mK}$, far below $-150\,{\rm mK}$. In contrast, fully filtered fields
  (2nd and 3rd rows) are shown in color schemes bounded by the actual
  minimum and maximum values.
\label{fig:UVonly}}
\end{figure*}

\begin{figure*}

\begin{center}

\includegraphics[width=0.90\textwidth]{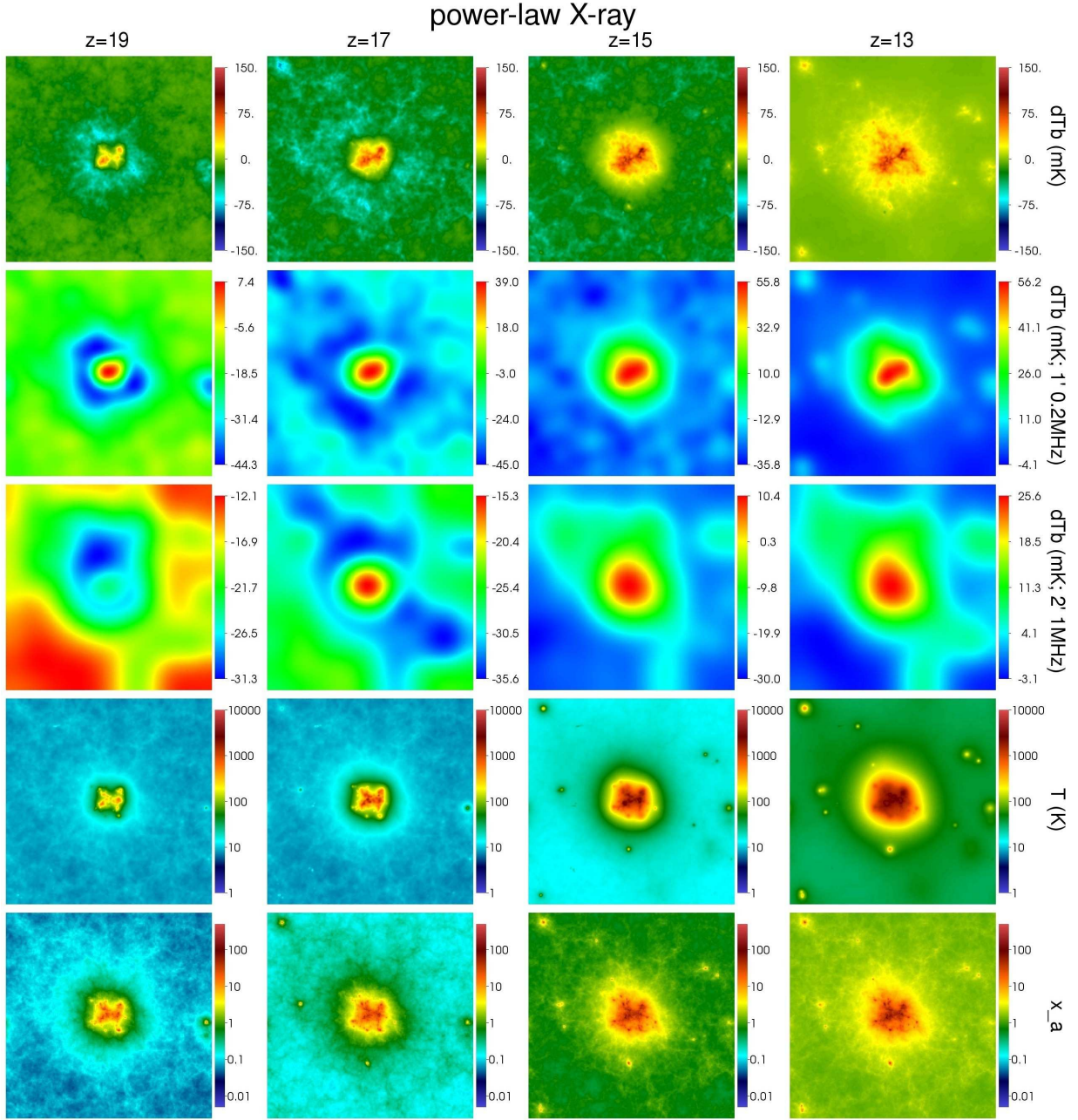}
\end{center}

\vspace{-0.2cm}
\caption{Same as Figure \ref{fig:UVonly}, but when the X-ray SED is
the composite power-law given 
by Equation~(\ref{eq:power-law-SED}).
\label{fig:synthetic}}
\end{figure*}

\begin{figure*}

\begin{center}

\includegraphics[width=0.90\textwidth]{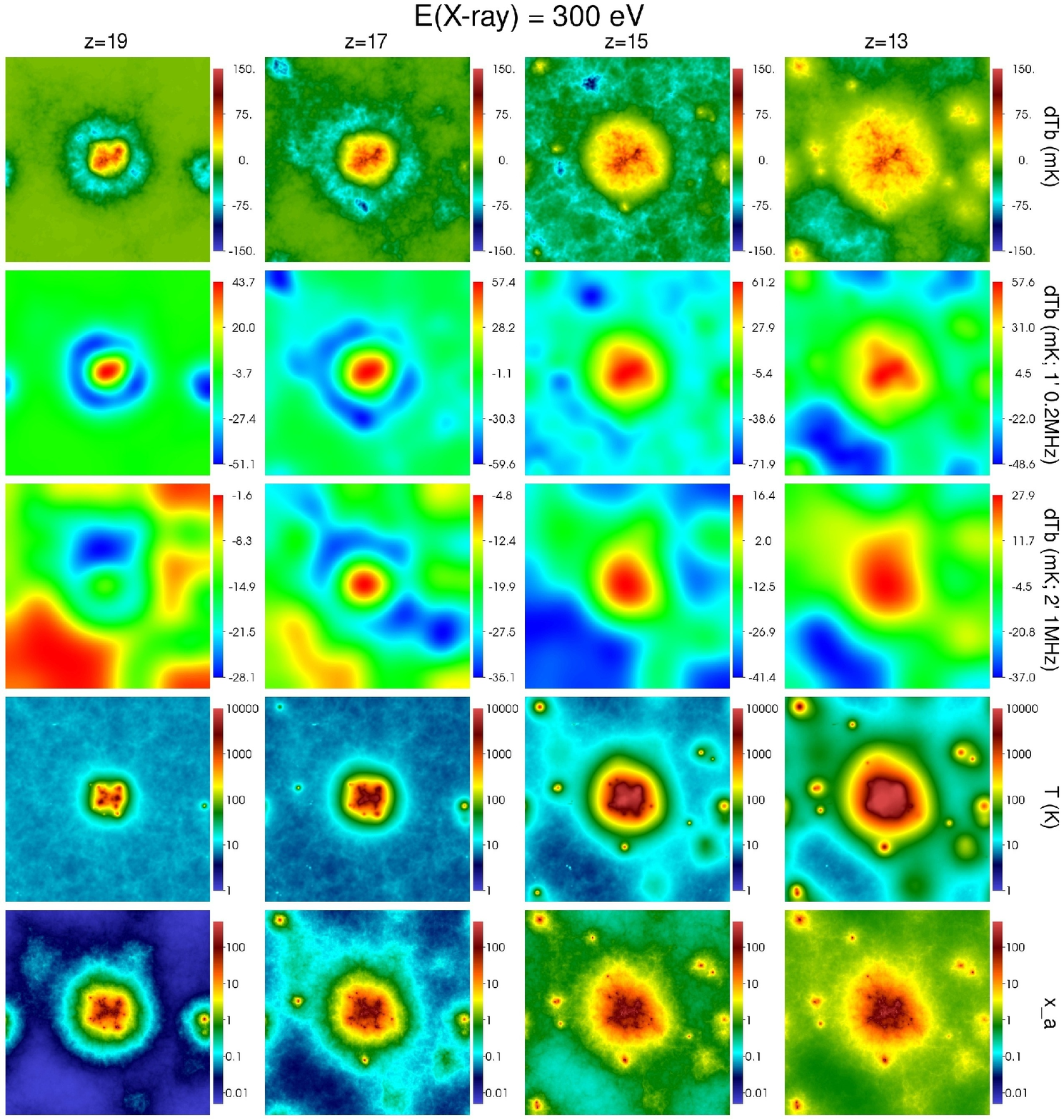}
\end{center}

\vspace{-0.2cm}
\caption{Same as Figure \ref{fig:UVonly}, but 
when $\epsilon_{{\rm X},0}=300$ eV.
\label{fig:300eV}}
\end{figure*}

The results at $z$ = 17, 15 and 13 on  either the $256^{3}$ real-space
grid or \#(sky pixels)$\times$\#(frequency bins)$=256^{2}\times 256$
real-sky+frequency-space grid
are illustrated in Figures \ref{fig:densvel}-\ref{fig:3keV}.
Even though we accurately treat the finite optical depth and the
effect of the peculiar velocity, it is still intuitive to analyze the result with the optically thin
approximation, under which
\begin{eqnarray}
\delta T_{b}&\simeq&35.1\,{\rm
  mK}\left(\frac{\Omega_{b}h^{2}}{0.0224}\right)\left(\frac{\Omega_{m}h^{2}}{0.135}\right)^{-0.5}
\nonumber \\
& & \times \left(\frac{T_{s}-T_{{\rm CMB}}}{T_{s}}\right)\left(\frac{1+z}{16}\right)^{0.5}\left(1+\delta_{{\rm HI}}\right)\left(1-x\right),\label{eq:deltaTb}
\end{eqnarray}
where $T_{{\rm CMB}}=2.725\,{\rm K}\,(1+z)$, 
$\delta_{{\rm HI}}=
({\rho_{{\rm HI}}-\bar{\rho}_{{\rm HI}}})/{\bar{\rho}_{{\rm HI}}}$,
and $x$ is the ionized fraction of hydrogen.
The overdensity of Rarepeak (Figure \ref{fig:densvel}) is on
average $\delta_{{\rm HI}}\sim0.64$, 
and is maximized at $\delta_{{\rm HI}}\sim17$ in this resolution,
which boosts $\delta T_{b}$ of the most heated region 
to $\sim35.1\,{\rm mK}\times[1.64-18]\sim[58-630]\,{\rm mK}$.
Source location, density, temperature, and coupling parameter ($x_{\alpha}$)
are rather strongly correlated with each other in the shock-heated
or ionized region. Thus, the dense inner region is
in saturated emission ($\delta T_{b}>0$) regime such that $T_{s}\sim T_{k}\gg T_{{\rm CMB}}$
and thus $\delta T_{b}=35.1\,{\rm mK}\left[(1+z)/16\right]^{0.5}\left(1+\delta_{{\rm HI}}\right)\left(1-x\right)$.
Outside Rarepeak, in contrast, $T_{k}<T_{{\rm CMB}}$ before the IGM is
fully heated ($z\gtrsim15$) and thus it is in absorption ($\delta T_{b}<0$).
The whole volume is seen in emission when (1) a large amount of X-ray
photons escape 
Rarepeak and (2) heat up gas efficiently, where the former occurs
more easily with harder SEDs and the latter with softer SEDs, resulting
in full emission at $z=13$ with photon energies in the range
$\epsilon_{{\rm X},\,0}=[0.5-1]\,{\rm keV}$. When $\epsilon_{{\rm X},\,0}=0.3\,{\rm keV}$,
X-ray photo-heating is most efficient, and thus the temperature
inside Rarepeak and other peaks are the highest, while efficient X-ray
photon trapping by all these peaks regulate the build-up of the X-ray
background in the IGM outside the peaks. Thus, a substantial portion
of the IGM remains colder than the CMB even at $z=13$. Because the X-ray
photon opacity and heating rate drop rapidly at high frequency ($\sigma(\nu)\sim\nu^{-3}$),
the $\epsilon_{{\rm X},\,0}=3\,{\rm keV}$ case has the smallest emission
region, while the IGM outside the shock-heated region is the coldest among
all cases and resembles the case without X-ray sources most. The
composite power-law SED case resembles the $\epsilon_{{\rm X},\,0}=770\,{\rm eV}$
case at $z=17$ and 15, while they show some difference at $z=13$.

The high efficiency of X-ray radiation from our assumed Pop-III X-ray
binaries in generating Ly$\alpha$ photons is clearly demonstrated in
Figures~\ref{fig:UVonly}-\ref{fig:3keV}. 
Except for the case with
  $\epsilon_{{\rm X},\,0}=3\,{\rm keV}$ (Figure~\ref{fig:3keV}),
the Ly$\alpha$ intensity (or $x_{\alpha}$) generated by secondary excitation
due to X-ray photons (Figures~\ref{fig:synthetic}-\ref{fig:1keV})
significantly exceeds the one generated 
by the UV continuum only (Figure~\ref{fig:UVonly}). 
This also indicates that X-ray binaries
are efficient sources of heating, as shown in Figures~\ref{fig:UVonly}-\ref{fig:1keV},
as long as most -- $50\%$ in our simulation -- Pop-III stars leave
behind X-ray binaries, but again with the exception of the case with
$\epsilon_{{\rm X},\,0}=3\,{\rm keV}$ (Figure~\ref{fig:3keV}). When
$\epsilon_{{\rm X},\,0}=3\,{\rm keV}$, the relatively small optical
depth makes the X-ray photons travel almost freely without too much
interaction with gas elements.

How emission transitions to absorption is also noteworthy, because
it shows a ``Ly$\alpha$ sphere''. When $\epsilon_{{\rm X},\,0}=3\,{\rm keV}$,
or there is no X-ray flux present, the emission region (heated by adiabatic
compression of the IGM) is surrounded by the maximum-absorption trough,
weakening towards larger radii (Figures \ref{fig:UVonly} and \ref{fig:3keV}).
This is due to the rapid decline of $x_{\alpha}$ ($N_{\alpha}$)
with increasing radius, which couples $T_{s}$ most strongly
to the $T_{k}$ ($<T_{{\rm CMB}}$) of the relatively unheated IGM in the
vicinity of clustered sources but progressively weakly outward. In
contrast, at $z\lesssim15$, in the cases of $\epsilon_{{\rm X},\,0}=[0.3-1]\,{\rm keV}$
and the composite power-law SEDs, $\delta T_{b}$ decreases monotonically with increasing radius. In these cases, X-ray heating
becomes very efficient and the IGM around the sources is heated significantly above
$T_{{\rm CMB}}$, so there is no chance to form the deep absorption
trough at $z\lesssim15$. Nevertheless, there still exist cases with
an absorption trough with X-rays present, which occur at $z\gtrsim17$
when the SED is a power law (at $z\sim19$) or $h\nu_{0}=0.3\,{\rm keV}$
(at $z\sim[17-19]$). This is due to the efficient X-ray photon trapping
at low frequency and the slower build-up of an X-ray background
than in the harder X-ray cases.

The strong emission, $\delta T_{b}\sim[200-630]\,{\rm mK}$, and the
absence of the absorption trough (except for the inefficient heating
case) in some of the models may seem to contradict previous predictions
on the 21-cm signal from individual QSOs or first star-black hole
systems (e.g. \citealt{Cen2006}; \citealt{2006ApJ...648L...1C};
\citealt{Chen2008}). The former is simply due to the fact that Rarepeak
has a large overdensity, while previous work assumed
a mean density for the IGM ($\delta_{{\rm HI}}=0$). The latter is due to the
fact that very efficient creation of X-ray binaries makes the
ratio of X-ray to UV photons larger than that used in previous work. Rarepeak
has $L_{X}\sim9\times10^{41}\exp[0.5(z-15)]\,{\rm erg\, s^{-1}\, SFR(M_{\odot}/yr)}$
(XAWNO; SFR=star formation rate) when both Pop III and II stars
are considered, while much smaller values, $L_{X}\lesssim5\times10^{40}\,{\rm erg\, s^{-1}\, SFR(M_{\odot}/yr)}$,
have been used previously to model starburst galaxies (e.g. \citealt{Chen2008}).
This can be cast into another familiar parameter $f_{X}$, defined
in \citet{2006MNRAS.371..867F} as 
\begin{equation}
L_{X}=3\times10^{40}\, f_{X}\,{\rm erg\, s^{-1}\, SFR(M_{\odot}/yr)}
\label{eq:fx}
\end{equation}
or sometimes more concisely as $f_{X}=\epsilon_{X}/(540\,{\rm eV})$
where $\epsilon_{X}$ is the total X-ray energy emitted per stellar
baryon(\citealt{2003MNRAS.340..210G}).
With the former definition, we find that $f_{X}\sim27$ at $z\sim15$
for efficient X-ray binary cases while X-rays from starburst galaxies
have $f_{X}\lesssim2$ at most. If we only consider the Pop III SFR, we
find an even larger value, $f_{X}({\rm Pop\, III})\sim15000$ at all
times. For the ``net'' $f_{X}$ (Pop III and Pop II together) to
be diluted to the value assumed for starburst galaxies, the SFR (dominated by Pop
II stars) should increase by an order of magnitude from $z=15$. 
This large value of $f_{X}({\rm Pop\, III})$ is mainly due to a very
optimistic scenario for the formation and evolution of the X-ray
binaries: black-hole formation rate out of binary stars $f_{\rm BH}\sim 50\%$,
long-duration ($t_{\rm acc}\sim 10-30\,{\rm Myr}$) and high-efficiency (Eddington
limit) accretion rate ($f_{\rm Edd}=1$). Also, we take the full X-ray bolometric luminosity
to estimate $f_X$, while \citet{2006MNRAS.371..867F} restrict the
energy range to $\gtrsim 2\,{\rm keV}$. All these effects add to
yield such a high $f_X$, compared 
to more typical values such as e.g. $f_X\sim 4$
when $f_{\rm BH}\sim 1\%$, $t_{\rm acc}\sim 20\,{\rm Myr}$, $f_{\rm
  Edd}=0.1$, and only $10\%$ of the total luminosity enters
Equation~(\ref{eq:fx}), as suggested by \citet{Mirabel2011}.

In previous work, because much smaller $f_{X}$ is used than ours, and consequently the
X-ray heating zone does not fully cover the Ly$\alpha$ sphere, it makes
the absorption trough always visible (\citealt{Cen2006}; \citealt{2006ApJ...648L...1C};
\citealt{Chen2008}; complete Ly$\alpha$ coupling, $T_{s}=T_{k}$,
is assumed in \citealt{Alvarez2010} and thus an absorption plateau,
rather than a trough, is predicted). 

In summary, a large variance in X-ray heating, X-ray trapping and Ly$\alpha$
pumping yield widely varying predictions for the 21 cm signal originating
from our simulation volume. The quasi-spherical absorption trough will
be a smoking gun for clustered high-redshift sources with strong UV
emission. The Pop III X-ray binaries, with high efficiency
in heating in most cases, form a new class of 21 cm profiles which
monotonically decrease outward from a concentrated region of strong
emission to either emission (late heating phase) or absorption plateau
(early heating phase), in addition to the cases with absorption troughs
when efficient X-ray trapping occurs. Detectability by existing and
future-generations of radio telescopes will be investigated in Section
\ref{sec:Detectability}. 

\section{Detectability of Rarepeak by SKA and SKA precursors}
\label{sec:Detectability}

\subsection{Tomography: Imaging individual peaks by SKA}
\label{sub:Tomography}

We investigate whether the strong contrast in $\delta T_{b}$ and
relatively large angular extension, $\theta\sim10'$, of Rarepeak
can be revealed through tomography by radio telescopes. SKA will,
after construction, have the highest sensitivity among all current
and planned radio telescopes, and thus we focus on SKA (SKA precursors
are indeed incapable of individual imaging of Rarepeak as shown here).
We have experimented with various combinations of the Gaussian beam
width $\Theta$ and frequency bandwidth $\Delta\nu$ under a reasonably
dedicated integration time, $T\gtrsim1000$ hr, and found that filters
of $\Theta\sim2'$ and $\Delta\nu\sim1\,{\rm MHz}$ can be considered
most optimal for tomography in most cases with reasonable resolution
and S/N ratio. In this section, we therefore report results based
on this specific filtering configuration, and also examine the impact
of a smaller filter of $\Theta\sim1'$ and $\Delta\nu\sim0.2\,{\rm MHz}$
for comparison.

\begin{figure*}

\begin{center} 

\includegraphics[width=0.90\textwidth]{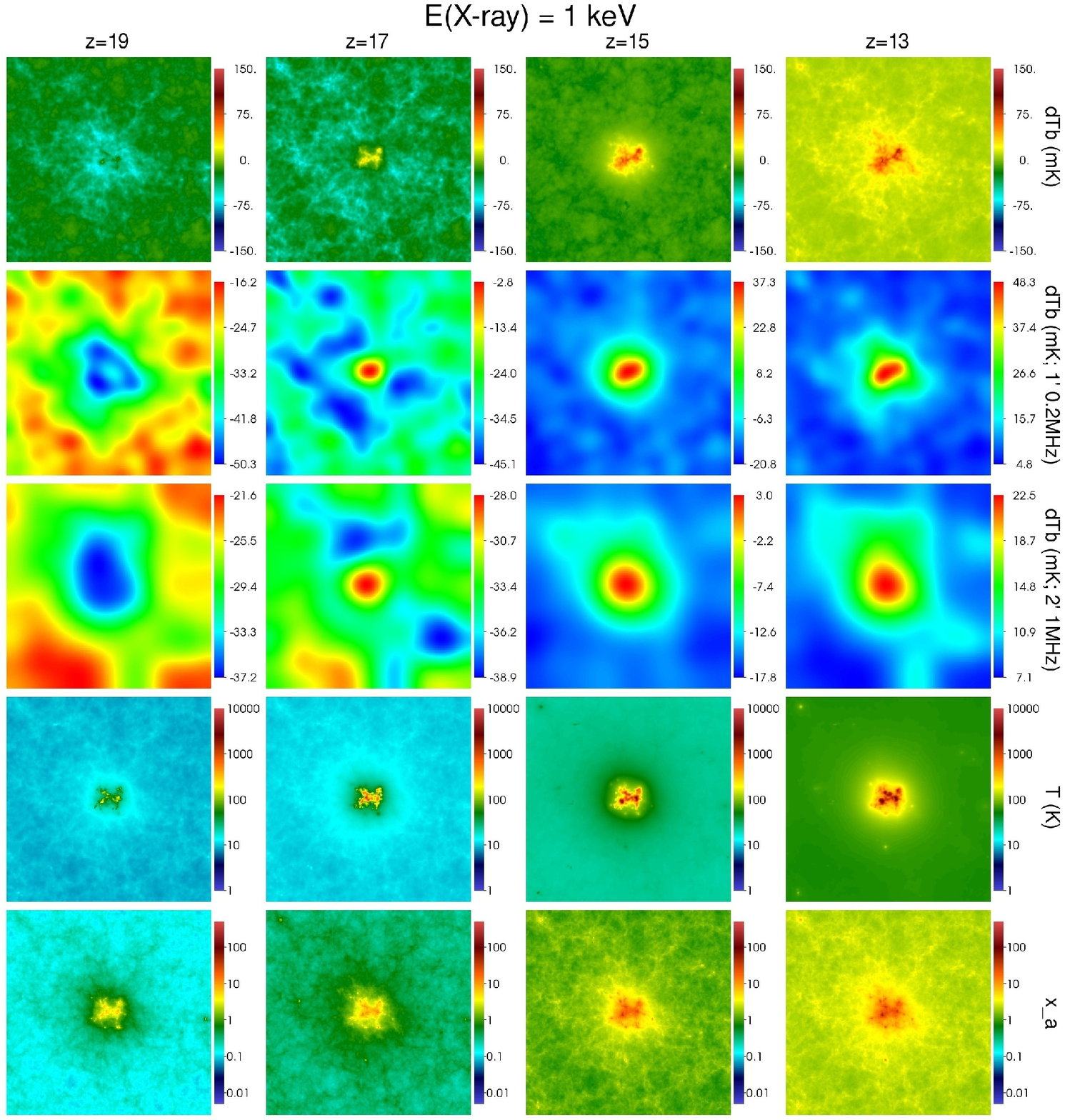}
\end{center}

\caption{ Same as Figure \ref{fig:UVonly}, but 
when $\epsilon_{{\rm X},0}=1$ keV.
\label{fig:1keV}}
\end{figure*}

\begin{figure*}
\begin{center}
\includegraphics[width=0.90\textwidth]{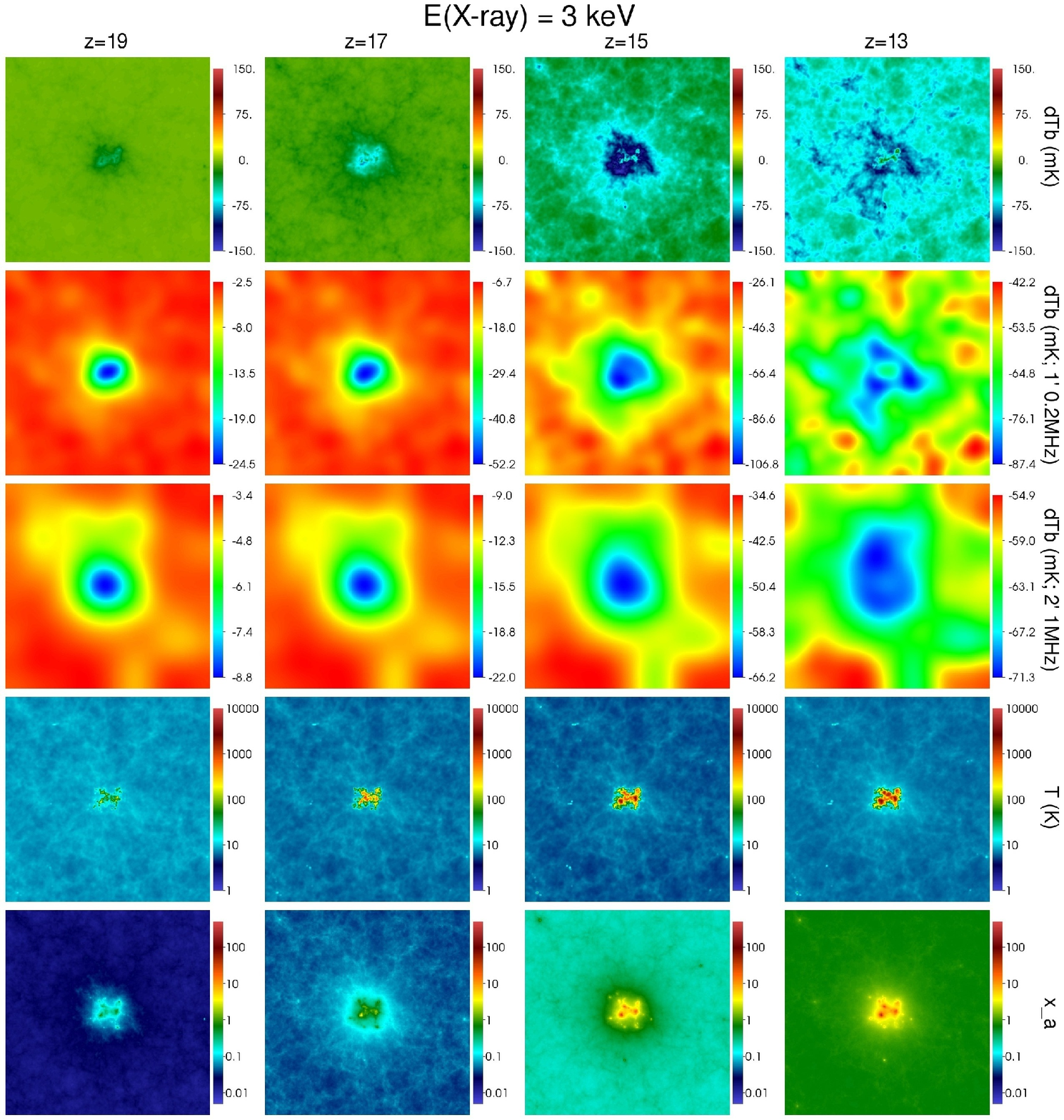}
\end{center}
\caption{ Same as Figure \ref{fig:UVonly}, but 
when $\epsilon_{{\rm X},0}=3$ keV.
\label{fig:3keV}}
\end{figure*}

In Figures \ref{fig:UVonly}-\ref{fig:3keV}, we
show the filtered $\delta T_{b}$ (denoted by $\widehat{\delta T_{b}}$
below) fields of the simulation box with 
$\{\Theta,\,\Delta\nu\}=\{1',\,0.2\,{\rm MHz}\}$
and $\{\Theta,\,\Delta\nu\}=\{2',\,1\,{\rm MHz}\}$. The central emission
feature is mostly erased inside the absorption region when a strong
absorption trough exists, while in low-energy X-ray cases the X-ray
heated region is extended just enough to be still discernible against
absorption with these filters ($h\nu_{0}=0.3\,{\rm keV}$ and
power-law SED cases at $z=19,\,17$). With the normalized ($\int^{\Delta\nu}d\nu\int d\theta^{2}R=1$)
response function of an instrument, approximated as 
\begin{equation}
R(\nu,\,{\bf \theta})=\frac{1}{2\pi\Theta^{2}\Delta\nu}{\rm e}^{-\theta^{2}/2\Theta^{2}},\label{eq:filter}
\end{equation}
the filtered signal at a real space pixel is given by
\begin{equation}
\widehat{\delta T_{b}}=\int_{\nu(z)-0.5\Delta\nu}^{\nu(z)+0.5\Delta\nu}\int_{0}^{\infty}d\theta^{2}\delta T_{b}(\nu,\,{\bf \theta})R(\nu,\,{\bf \theta}),\label{eq:dtb_hat}
\end{equation}
where the angle $\theta$ is measured from the beam center. The real-space
pixel noise, after integration time $t$, is given by (see e.g. \citealt{Chen2008})
\begin{equation}
T_{{\rm N}}\simeq\frac{T_{{\rm sys}}}{f_{{\rm cov}}\sqrt{\Delta\nu\, t}}\simeq\frac{1000\,{\rm K}\left(\frac{1+z_{s}}{16}\right)^{2.5}}{f_{{\rm cov}}\sqrt{\Delta\nu\, t}},\label{eq:noise}
\end{equation}
where the system temperature $T_{{\rm sys}}$ is assumed to be dominated
by the galactic synchrotron background, and the covering factor $f_{{\rm cov}}\equiv{N_{{\rm dish}}A_{{\rm dish}}}/{A_{{\rm tot}}}$
is the fraction of the land area $A_{{\rm tot}}$ covered by $N_{{\rm dish}}$
dishes (or stations) of the individual area $A_{{\rm dish}}$, which
also corresponds to the Fourier-space coverage fraction. Then
the S/N ratio becomes
\begin{eqnarray}
{\rm S/N}=\frac{\widehat{\delta T_{b}}}{T_{{\rm N}}}&\simeq& 19f_{{\rm
    cov}}\left(\frac{1+z_{s}}{16}\right)^{-2.5} \nonumber \\ 
&&\times\left(\frac{\Delta\nu\, t}{1\,{\rm MHz\,1000\,{\rm
      hr}}}\right)^{1/2}\left(\frac{\widehat{\delta T_{b}}}{10\,{\rm
    mK}}\right).
\label{eq:sn_ratio}
\end{eqnarray}

For an instrument like SKA, which is composed of a core (or cores)
with densely spaced stations and wings with sparsely spaced stations,
$f_{{\rm cov}}$ should be defined differently because $A_{{\rm tot}}\sim B^{2}$
, where $B$ is the baseline corresponding to the intended angular
resolution $\Theta$. We therefore simply adopt the already calculated
S/N ratio at $z=15$ under a specific configuration from \citet{Mellema2013},
and assume that this fixes $f_{{\rm cov}}$ at $z=15$. For example,
with $\Theta=2'$, $\Delta\nu=1\,{\rm MHz}$ and $t=1000\,{\rm hr}$,
SKA with the total collecting area of $1\,{\rm km^{2}}$, the core
of $2\,{\rm km}$-diameter with the baseline distribution of (distance)$^{-1}$
within the core gives ${\rm S/N}=\left({\widehat{\delta T_{b}}}/{10\,{\rm mK}}\right)$
at $z=15$ (Figure 23 of \citealt{Mellema2013}). If we assume that
$N_{{\rm dish}}$ increases very slowly when the baseline increases
beyond the core size, $f_{{\rm cov}}\propto1/A_{{\rm tot}}\propto1/B^{2}\propto(1+z)^{-2}\Theta^{2}$,
because $B=\lambda/\Theta=5.8\,{\rm km}\left[({1+z_{s}})/{16}\right]({\Theta}/{2'})^{-1}$.
With this we find that this specific SKA configuration roughly gives
\begin{eqnarray}
\left[{\rm S/N}\right]_{{\rm
    2km-core-SKA}}&\simeq&\left(\frac{1+z_{s}}{16}\right)^{-4.5}\left(\frac{\Theta}{2'}\right)^{2}\left(\frac{\widehat{\delta T_{b}}}{10\,{\rm mK}}\right)
\nonumber \\
& &\times \left(\frac{\Delta\nu\, t}{1\,{\rm MHz\,1000\,{\rm hr}}}\right)^{1/2}.\label{eq:ska_sn_ratio}
\end{eqnarray}
The noise of SKA to real space pixel is then $\sim$27, 17, 10, and
5 mK at source redshifts $z_{s}=$19, 17, 15, and 13, respectively,
with $\Theta\sim2'$, $\Delta\nu\sim1\,{\rm MHz}$ for $t=1000$ hr
integration. With $\Theta\sim1'$, $\Delta\nu\sim0.2\,{\rm MHz}$
for $t=1000$ hr integration, the sensitivity becomes $\sim9$
multiples of the above.

\begin{figure}[ht]
\includegraphics[width=0.5\textwidth]{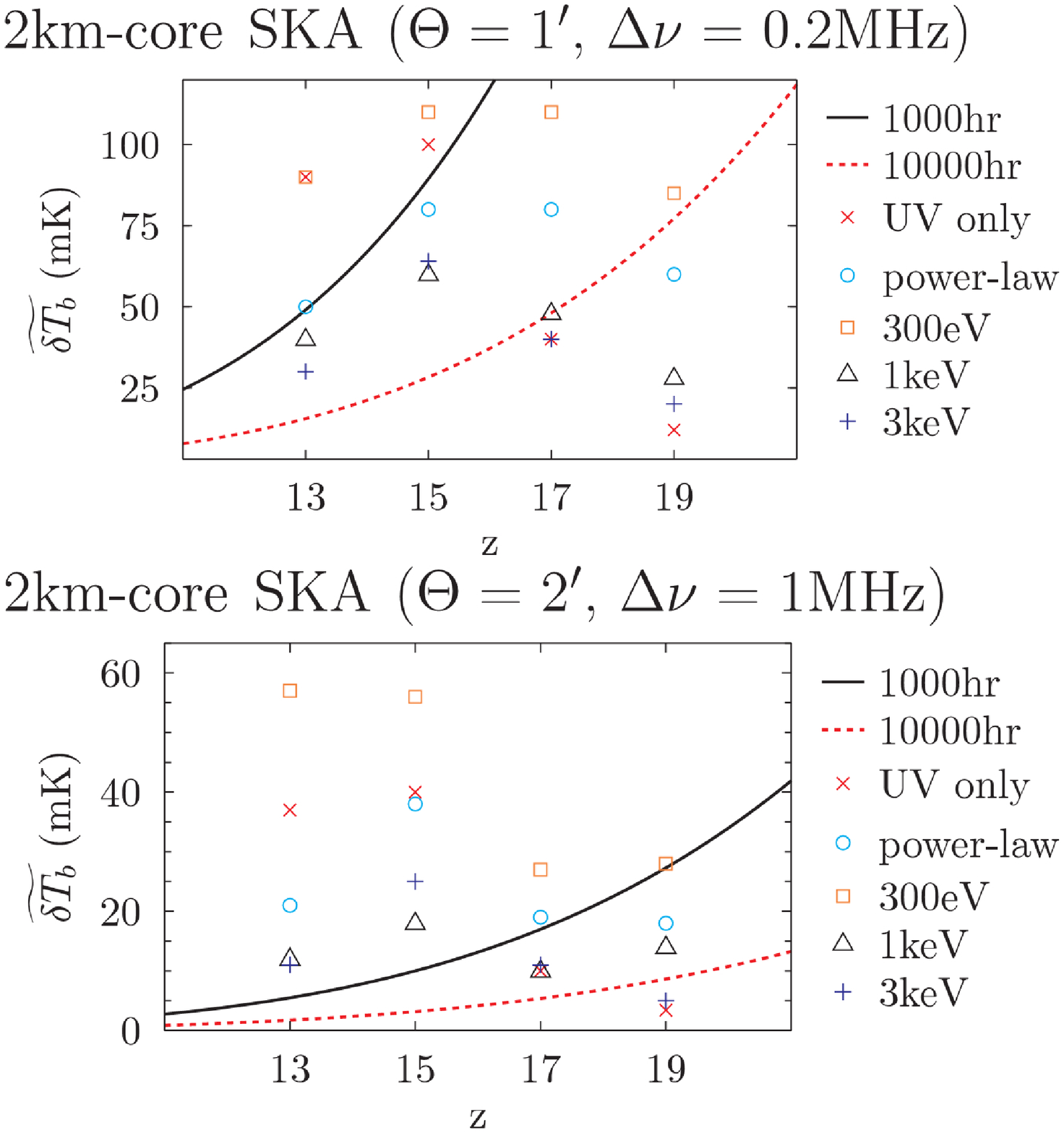}

\caption{Detectability of Rarepeak by SKA. Two different filter configurations
(angular resolution and bandwidth) are chosen, and the signals ($\widetilde{\delta T_{b}}$)
are shown against the noises with different integration times (black
curve:1000 hr; red curve: 10000 hr). The top panel and the bottom
panel are in conjuction
with the 2nd and 3rd rows of Figures~\ref{fig:UVonly}-\ref{fig:3keV},
respectively.  Note that identical values of
$\widetilde{\delta T_{b}}$ 
do not always mean identical features (e.g. in the top panel, at
$z=13$, UV-only and 300 eV cases have the same strength in signal
but the former is in absorption while the latter is in emission).
\label{fig:imaging-detectability}}
\end{figure}

Since mean separation between similar rare density peaks is expected
to be large ($\sim 100\,{\rm Mpc}$, Section \ref{sub:Power-spectrum-analysis}), these peaks
will be seen as sparse islands against the mean. Thus the proper ``deviation''
of $\widehat{\delta T_{b}}$, $\widetilde{\delta T_{b}}$, is roughly
the difference between the filtered value at the peak center, $\widehat{\delta T}_{b,\,{\rm center}}$,
and the minimum value, $\widehat{\delta T}_{b,\,{\rm min}}$, outside.
The detectability is depicted in Figure \ref{fig:imaging-detectability}.
As shown, 2-km core SKA will be able to image most cases at high S/N
ratio at $z\lesssim15$, and low-energy cases (300 eV and power-law
SED) can be barely seen at ${\rm S/N}\sim1-1.5$ at $z\sim17$, under
$\{\Theta,\,\Delta\nu,\, t\}=\{2',\,1\,{\rm MHz},\,1000\,{\rm hr}\}$.
To achieve similar detectability with better resolution $\{\Theta,\,\Delta\nu\}=\{1',\,0.2\,{\rm MHz}\}$,
one needs $t\sim10000\,{\rm hr}$, where the increase in $t$ is somewhat
compensated by the increase in $\widetilde{\delta T_{b}}$. It is
obvious that imaging will become much more difficult with SKA precursors,
when $\widetilde{\delta T_{b}}\sim[10-60]\,{\rm mK}$ (under $\{\Theta,\,\Delta\nu\}=\{2',\,1\,{\rm MHz}\}$)
at $13\lesssim z\lesssim17$. 

We can extrapolate our result to generic cases. If X-ray heating is
either minimal or trapped in the immediate vicinity of the sources, imaging
individual clusters of sources would become possible even with the
precursors at lower redshifts. In contrast, if efficient heating occurs
too early, it will become more difficult to observe these peaks through
imaging by the precursors.

We therefore conclude that 2-km core configuration for SKA (\citealt{Mellema2013})
is preferred for probing astrophysical objects at $z\sim17$. At slightly
lower redshifts, the deep-absorption feature would be detectable at
high S/N ratio. Because this feature is not possible otherwise, 
it will be a smoking gun for clustered radiation sources of strong
UV emission at high redshifts when the ionization state of the Universe
is still very low.

\subsection{Power spectrum analysis: Detecting many rare peaks
  statistically}
\label{sub:Power-spectrum-analysis}

While only SKA is usable for imaging radiative structures of individual
peaks at $z\sim13-17$ (Section \ref{sub:Tomography}), one may ask
whether multiple peaks can generate 21-cm signal detectable by lower-sensitivity
apparatuses through power spectrum analysis. We therefore investigate
this possibility by calculating the power spectrum due to these rare
peaks. This is to
quantify the contribution to the power spectrum from these {\em rare peaks
only}, and checking for any possibility to sort these high-density
peaks out of
other contributors such as the fluctuation of the cosmological matter-density field
and the fluctuation in the distribution of all the astrophysical sources.
In reality, if all the astrophysical sources are considered,
it is usually found that the
amplitude of the power spectrum is  dominated by these astrophysical
sources, ranging $1-2$ orders of magnitude larger than that of the
cosmological origin (e.g. \citealt{Pritchard2007}, \citealt{Baek2010} and
\citealt{Pacucci2014}). Therefore, if the resulting power spectrum is
comparable to the cosmological one, we may safely assume that the
power spectrum analysis is not adequate for probing these rare objects.
Note that this is a rough estimate of the detectability, and
thus in this paper we simply take the real-space 3D power spectrum
in the comoving coordinates as the approximate 3D power spectrum in
the observing frame, where one should in principle take into account
the impact of the peculiar velocities (e.g. \citealt{2005ApJ...624L..65B};
\citealt{2006ApJ...653..815M}; MSMIKA) and the light-cone
effect (e.g. \citealt{Datta2012a}; \citealt{Zawada2014}).

In a Gaussian density field filtered on scale $R_{w}$,
the number density of peaks with $\delta>\nu\sigma_{R_{w}}$ ($\sigma_{R_{w}}$:
rms of filtered density) is given by (\citealt{Kaiser1985})
\begin{equation}
n_{{\rm peak}}(>\nu)\simeq\frac{1}{4\pi^{2}}\left(\frac{n+3}{6R_{w}^{2}}\right)^{3/2}\left(\nu^{2}-1\right)\exp\left(\frac{-\nu^{2}}{2}\right),\label{eq:peak-number-density}
\end{equation}
where $n$ is the power-law spectral index of matter-density power
spectrum $P_{{\rm \rho\rho}}(k)$. At $z\sim15$, around relevant
wavenumber $k={2\pi}/{R_{w}}\sim{2\pi}/{3\,{\rm Mpc}}\sim2\,{\rm Mpc^{-1}}$,
$P_{\rho\rho}(k)\propto k^{-2.3}$. The refined region has $\nu\sim3.45$
(where we assign filtering scale such that $({4\pi}/{3})R_{w}^{3}=${[}Eulerian
volume of refined region{]}). We obtain $n_{{\rm peak}}(>\nu=3.45)\sim1.1\times10^{-6}\,{\rm Mpc^{-3}}$
at $z\sim15$, and the mean separation between similar peaks $l_{{\rm peak}}\sim98.1\,$Mpc
(comoving).

\begin{figure*}
\includegraphics[width=1\textwidth]{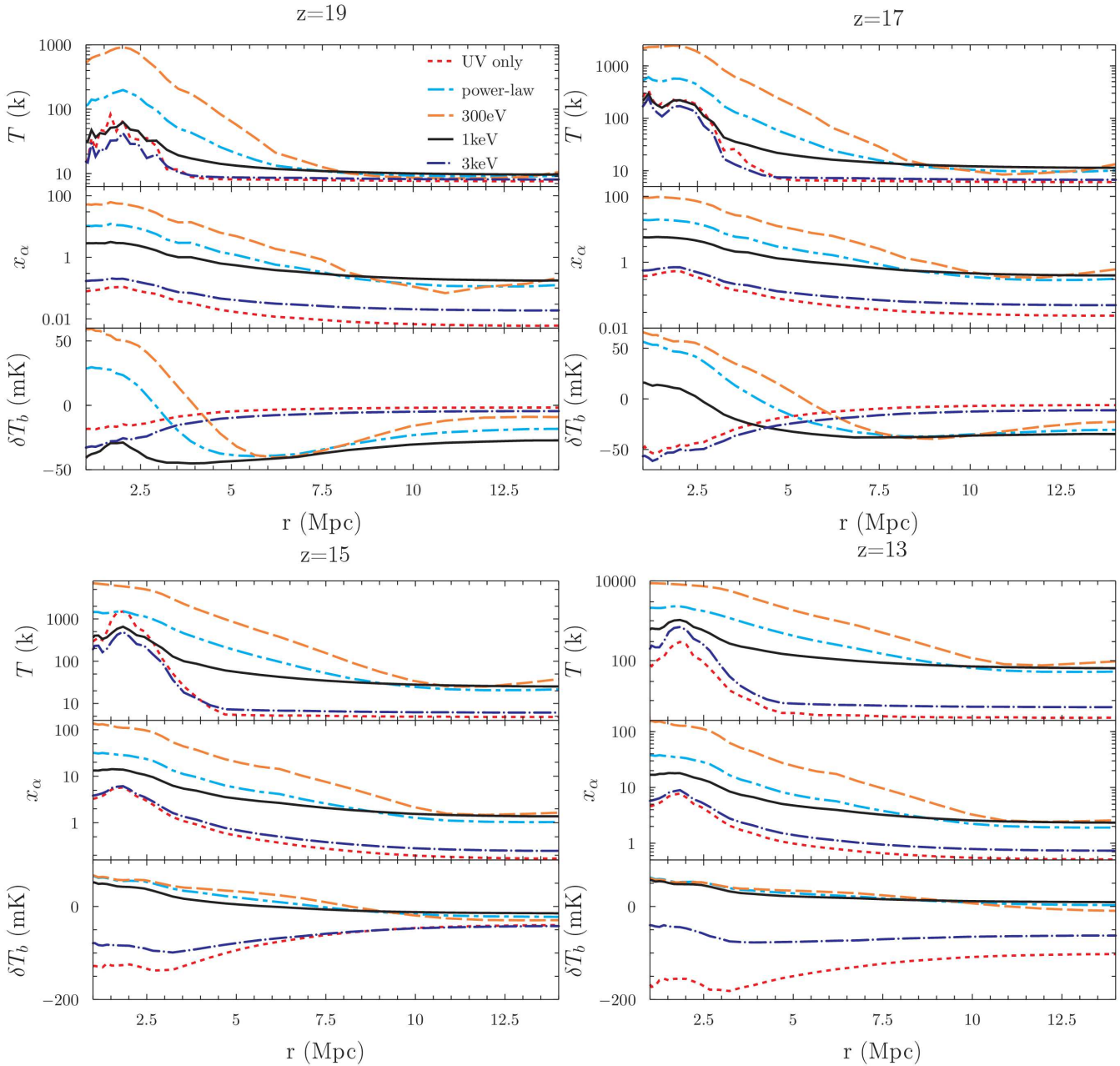}

\caption{Radially averaged profiles of $T$, $x_{\alpha}$ and $\delta
  T_{b}$, with $r$ 
measured from the center of Rarepeak in units of comoving Mpc.
Note that the Ly$\alpha$ sphere is seen in various forms of
  $\delta T_{b}$
  depending upon the model and the redshift.
\label{fig:profile}}

\end{figure*}

Let us now roughly estimate the expected power spectrum of $\delta T_{b}$
from these rare peaks with $\nu\gtrsim 3.45$ but ignoring all the other peaks.
We follow the convention that $\left\langle \delta T_{b}({\bf x})\delta T_{b}({\bf x}+{\bf r})\right\rangle =\int({d^{3}k}/{(2\pi)^{3}})P(k)\exp(-i{\bf k}\cdot{\bf r})$
in defining $P(k)$ of $\delta T_{b}$: we do not consider the monopole
term $P(k=0)$ here, which is not detectable by radio interferometers,
and thus $\delta T_{b}({\bf x})$ can be replaced by $(\delta T_{b}({\bf x})-{\rm constant})$
in the definition of $P(k)$ without affecting $P(k\neq0)$. We first
approximate $\delta T_{b}({\bf x})$ of each peak by a spherically
symmetric profile $\delta T_{b}(r)$, which is the radially averaged
$\delta T_{b}({\bf x})$ around the center of Rarepeak
(Figure~\ref{fig:profile}). We also assume 
for simplicity that all the peaks with $\nu>3.45$ have identical
profile: $\delta T_{b}(r;\,\nu>3.45)=\delta T_{b}(r;\,\nu=3.45)$.
Rarer (so higher-density) peaks will most likely have different profiles,
but as their abundance decreases quickly, this assumption does not
affect the answer much. If we then assume a random distribution of such
identical peaks with a given $n_{{\rm peak}}$, we can write the 3D
field of $\delta T_{b}$ as a sum of individual profiles centered
at $\left\{ {\bf x}_{i}\right\} $:
\begin{equation}
\delta T_{b}({\bf x})=\sum_{i}u(\left|{\bf x}-{\bf x}_{i}\right|)+\delta T_{b,\,{\rm IGM}},\label{eq:dtb-sum}
\end{equation}
where $u(r)\equiv\delta T_{b}(r)-\delta T_{b,\,{\rm IGM}}$. Here
we have assumed that $\delta T_{b}(r)$ flattens out beyond some cut-off
radius $r_{c}$, such that $\delta T_{b,\,{\rm IGM}}\equiv\delta T_{b}(r>r_{c})=\delta T_{b}(r=r_{c})$=constant.
This then enables the simple halo approach in calculating the nonlinear
power spectrum (e.g. \citealt{Loeb2013}). We first expect a shot
noise term, but not completely white -- meaning $P(k)$=constant --
due to the extended radial profile:
\begin{equation}
P_{{\rm 1h}}(k)=n_{{\rm peak}}u^{2}(k),\label{eq:1halo}
\end{equation}
where
\begin{equation}
u(k)\equiv\int_{0}^{r_{c}}dr\,4\pi r^{2}\frac{\sin(kr)}{kr}u(r)\label{eq:uk}
\end{equation}
is the Fourier transform of $u(r)$. $P_{{\rm 1h}}(k)$ is the usual
1-halo term, which reflects both the rarity of peaks ($n_{{\rm peak}}$)
and the characteristics of the radial profile. When these peaks are
spatially correlated, the 2-halo term
\begin{equation}
P_{{\rm 2h}}(k)=n_{{\rm peak}}^{2}u(k)^{2}b^{2}P_{\rho\rho}(k)\label{eq:2halo}
\end{equation}
arises with the bias parameter $b$, and thus $P(k)=P_{{\rm 1h}}(k)+P_{{\rm 2h}}(k)$.
If we simply use the linear bias parameter for $b$, then $b=1+(\nu^{2}-1)D(z)/1.68646$
where $D(z)$ is the growth factor in $\Lambda$CDM (\citealt{Mo1996}).
We find that for these peaks, $P_{{\rm 2h}}(k)\ll P_{{\rm 1h}}(k)$
at any $k$ and $z$ in all the models, which is a reflection of the
weak spatial correlation between rare objects. The variance at given
$k$, $\Delta^{2}(k)\equiv({1}/{2\pi^{2}})k^{3}P(k)$, is a useful
indicator of the degree of fluctuation at the corresponding scale.

\begin{figure*}
\includegraphics[width=1\textwidth]{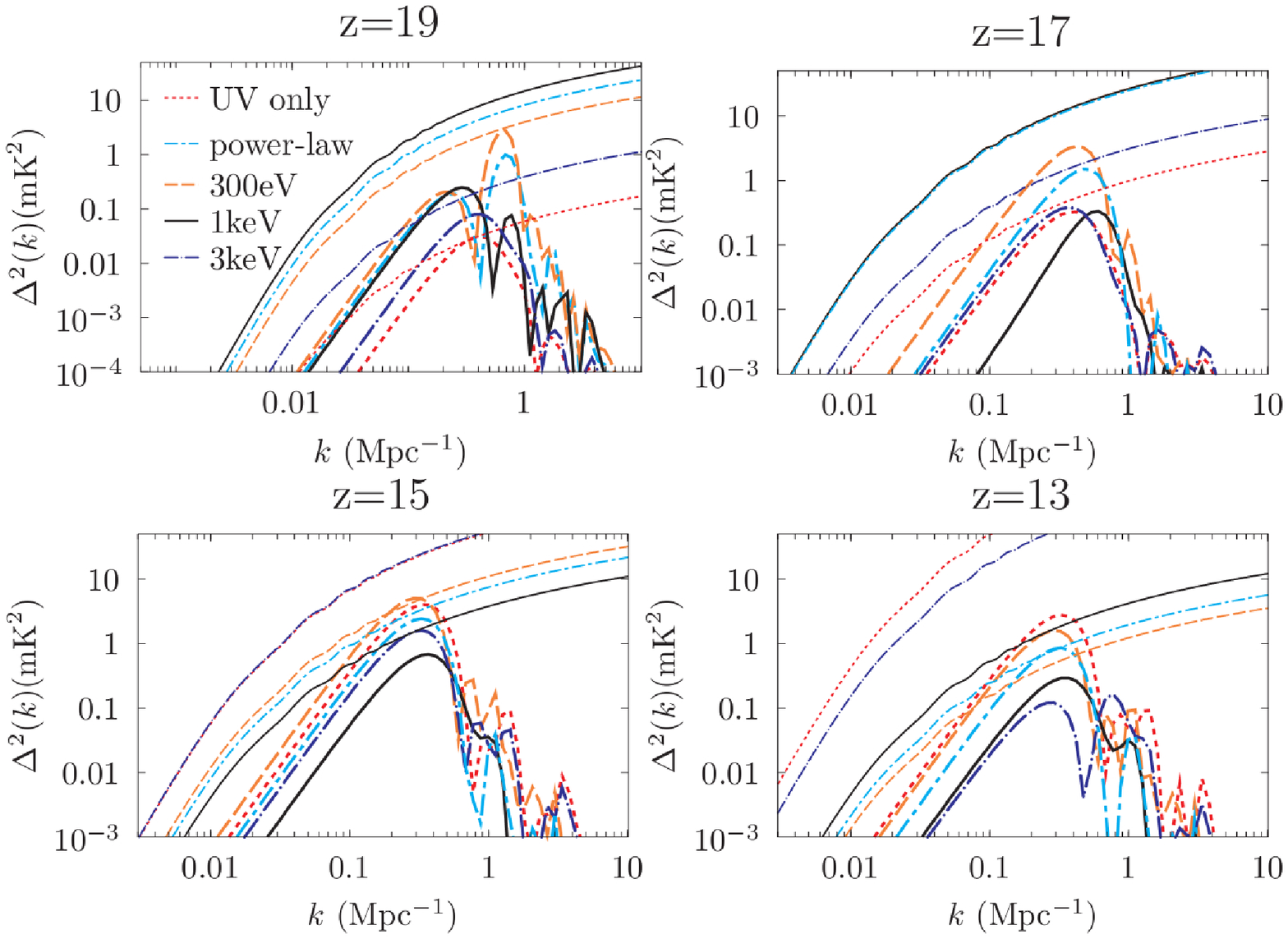}

\caption{21 cm power spectrum ($\Delta^{2}(k)=\frac{1}{2\pi^{2}}k^{3}P(k)$)
of rare peaks (thick curves) with $\nu\gtrsim3.5$ (at the filtering
scale $R_{w}\sim3\,{\rm Mpc}$). We also plot the 21 cm power
spectrum originating from linear density fluctuations, $\frac{1}{2\pi^{2}}k^{3}P_{{\rm c}}(k)=\frac{1}{2\pi^{2}}k^{3}P_{\rho\rho}(k)\delta T_{b,\,{\rm IGM}}^{2}$,
to see whether rare peaks can be distinguished in the power spectrum
analysis. Note that $\delta T_{b,\,{\rm IGM}}$ and $P_{{\rm c}}(k)$
are dependent on the X-ray model.
\label{fig:power-spectrum}}
\end{figure*}

As seen in Figure \ref{fig:power-spectrum}, $\Delta^{2}(k)$ of rare
peaks never
reaches $10\,{\rm mK}^{2}$ regardless of the X-ray model, which makes
the statistical observation of these peaks very difficult. The shape
of the power spectrum is given by (1) the white spectrum at $k\lesssim0.3-0.5\,{\rm Mpc}^{-1}$
because peaks are not resolved and are therefore equivalent to randomly distributed point sources on
these large scales, and (2) the spectrum roughly decaying faster than
$\sim k^{-3}$ at $k\gtrsim0.5\,{\rm Mpc}^{-1}$ due to the relatively smooth spatial
variation of $\delta T_{b}(r)$. The peak is at $k\sim 0.3\,{\rm
  Mpc}^{-1}$, and if we roughly extrapolate the noise estimates
at $k=0.1\,{\rm Mpc}^{-1}$ and $k=0.2\,{\rm Mpc}^{-1}$ with
$t=2000\,{\rm hr}$ of
\citet{Mesinger2014} to $k=0.3\,{\rm Mpc}^{-1}$, HERA and SKA will have
$\Delta^{2}(k)=9\,{\rm mK}^2 (t/2000\,{\rm hr})^{-1}$  and
$\Delta^{2}(k)=0.6\,{\rm mK}^2 (t/2000\,{\rm hr})^{-1}$,
respectively. Therefore, HERA and SKA may seem to have opportunity to observe
these rare peaks through the power spectrum analysis. However, as
noted earlier in this section, the net contribution from all the
astrophysical sources dominates the power spectrum and is much larger
than the power spectrum of the cosmological origin. Therefore, we need
to address whether the power spectrum only by these rare peaks is
comparable to the cosmological one.

We compare the nonlinear
$P(k)$ from rare peaks to the cosmological $P(k)$, or
\begin{equation}
P_{{\rm c}}(k)=\delta T_{b,\,{\rm IGM}}^{2}P_{\rho\rho}(k),
\label{eq:Pk-cosmological}
\end{equation}
where for consistency we use the IGM value $\delta T_{b,\,{\rm IGM}}$
which varies over different X-ray models, and assume zero ionization,
uniform background, uniform $T_{s}$ and take baryons as a perfect
tracer of dark matter\footnote{Note that baryons and dark matter particles have nonzero relative
bulk velocity -- denoted by $v_{{\rm bc}}$ -- as shown by \citet{Tseliakhovich2010},
and this can change the amplitude and the shape of $P_{{\rm c}}(k)$
significantly from the form given by Equation (\ref{eq:Pk-cosmological})
if this effect results in efficient thermalization of the IGM (\citet{McQuinn2012}).
We here restrict ourselves to the cases of relatively inefficient
thermalization, or $\lesssim0.33\,\%$ thermalization of $v_{{\rm bc}}^{2}$,
such that $P_{{\rm c}}(k)$ is given by Equation (\ref{eq:Pk-cosmological}),
and leave the investigation on how $v_{{\rm bc}}$ could change our
answer (e.g. impact on star formation efficiency, background build-up
and heating) for future work.}, such that $\delta T_{b}({\bf x})=\left\langle \delta T_{b}\right\rangle {\rho({\bf x})}/{\bar{\rho}}$. 
As seen in Figure \ref{fig:power-spectrum}, in most cases the power
spectrum is dominated by $P_{{\rm c}}(k)$. Nevertheless, there are
a few cases when $P(k)\gtrsim P_{{\rm c}}(k)$, occurring most prominently
in the case of 300 keV X-ray at $z\sim13$. For power-law X-ray SED
case, $P(k)\sim P_{{\rm c}}(k)$ at $z\lesssim15$.
Because we find
that this rare-peak-only power spectrum is comparable to the
cosmological one, we conclude that probing these rare peaks is
practically impossible with the power spectrum analysis, even with
HERA and SKA. Note that the peculiar shape of the power spectrum by
rare peaks (only) is the result of considering only those sparsely
spaced but individually extended signals, just as seen in similar
shape found by \citet{Alvarez2010} when only isolated QSO signals are
considered. In contrast, a very different shape is usually predicted
for the power spectrum when all astrophysical sources are considered,
which will be distributed  diffusely when they are strongly
correlated with the underlying
matter-density distribution
(e.g. \citealt{Pritchard2007}).

Based on the fact that the power spectrum from all the astrophysical
sources dominates the power spectrum over that of the cosmological
origin, and the power spectrum of rare peaks are just comparable to
the cosmological one, we conclude that probing these rare peaks through power
spectrum analysis is not as encouraging as tomography. 
Nevertheless, it is worthwhile to investigate how the detailed
microphysical process including various feedback effects may affect
the net power spectrum. Previous studies usually consider simplified
source models with an averaged or a constant mass-to-light ratio, which
makes the distribution of astrophysical sources very strongly
correlated with the 
underlying density field (e.g. \citealt{Pritchard2007},
\citealt{Baek2010} and \citealt{Pacucci2014}). If e.g. radiative
feedback effects affect star formation inside nearby halos, such a
strong correlation may break down and the power spectrum will be
affected. Estimating this self-consistently requires extending
  the degree of accuracy of our 
Rarepeak calculation to a much larger scale, which is beyond the scope
of this paper. We delay this investigation into the future (Xu et
al. in preparation).

\section{Summary and Discussion}
\label{discussion}

We have investigated the possibility of observing strongly clustered
stars at high redshifts ($z\sim13-19$) by future and current radio
interferometers through 21-cm tomography and power spectrum analysis.
We have assumed that X-ray emission is from X-ray binaries after binary
Pop III stellar systems evolve, with high success rate of 50\%
for such events (we let 50\% of Pop III star particles in the simulation end
up as X-ray binaries, and emit X-rays constantly for 10 Myrs
through the accretion of matter by black holes at the Eddington limit).
By properly calculating the local feedback and the
global background, these X-ray sources are found to be not efficient
in ionizing the IGM at $z\gtrsim13$, but very efficient in heating
the IGM above CMB temperature and also generating Ly$\alpha$ photons
when the rest-frame X-ray photon energy is $\lesssim1\,{\rm keV}$.
There exist somewhat isolated rare density peaks, which then generate
spatially extended ($\sim10$ comoving Mpc), quasi-spherical $\delta T_{b}$
profiles. The peak center is in emission, and depending on the X-ray
SED and its amplitude, there may or may not exist an absorption trough.
Low-energy X-rays are found to be very efficient in heating the IGM
before Ly$\alpha$ pumping becomes important ($x_{\alpha}\gtrsim1$),
for this high success rate of X-ray binary formation by Pop III stars,
erasing the absorption trough feature. When the rest-frame X-ray
energy is $\lesssim300\,{\rm eV}$, there exists an era when Ly$\alpha$
pumping is efficient while the X-ray heating region is somewhat confined
within the Ly$\alpha$ sphere in the cold IGM, leaving the absorption
trough still visible. Real space imaging (tomography) seems more
promising than the power spectrum analysis, and SKA with $\sim2\,{\rm km}$
core and reasonable (\textasciitilde{}1000
hr) integration times can obtain excellent S/N ratios for most of our models in this high redshift range, provided that 
the image plane is filtered with a beam of $\sim2'$ and bandwidth of
$\sim1\,{\rm MHz}$. The power spectrum due to many rare peaks such
as the ones treated here is smaller than that from linear fluctuations
in most cases, and thus the 21-cm power spectrum analysis is not likely
to probe these rare peaks out of the full 21-cm power spectrum
caused both by the cosmological density fluctuation and that by all
existing astrophysical sources.

We note that the absorption trough feature will be a smoking gun for
high-redshift UV sources. The depth of the absorption will strongly
depend on the efficiency of IGM heating. If the X-ray heating era
is more delayed because formation of the X-ray binary systems is not
as efficient as we assumed or rest-frame X-ray photons are predominantly
at low energies ($\lesssim100\,{\rm eV}$) such that they are mostly
trapped by the surrounding IGM, even deeper (a few to several hundred
mK when unfiltered) absorption will be observed. Nevertheless, as
the Universe evolves, more and more density peaks and filaments will
appear around Rarepeak, which may erase this isolated, quasi-spherical
feature by forming their own radiation profiles and overlapping with
one another (as seen in Fig \ref{fig:300eV}). We will investigate
this possibility in the future.

If an object like Rarepeak is imaged in the sky, it should reveal the degree
of X-ray heating from the system. If no absorption trough is observed,
it will imply that at least locally the IGM has been heated beyond
the CMB temperature (or simply the trough is unresolved). In contrast,
when an absorption trough is observed, it will imply that either
Ly$\alpha$ pumping efficiency (and hence the UV luminosity) is much
greater than the X-ray heating efficiency (X-ray luminosity) or the
system is in the early stage of X-ray heating such that the heated
region has not grown comparable in size to the Ly$\alpha$ sphere. In either case,
we should be able to constrain the UV/X-ray luminosity ratio of these
early sources. Depending on the aggregated UV/X-ray luminosity ratio
from the full spectrum of sources, it may be possible that the X-ray
heating era can precede the Ly$\alpha$ pumping era, while we do not
observe this in any of our models.

There seems to be tension between 21-cm astrophysics and cosmology even at
high redshifts ($z\gtrsim 15$).
While it is difficult to extrapolate our result directly, there exist
several astrophysical effects that require careful attention for
high-redshift 21-cm cosmology. First,
it may be possible that many well isolated peaks produce 
nonlinear fluctuations in the Ly$\alpha$ and the X-ray background,
as it seems already possible at a specific $k$ range with an admittedly
contrived X-ray model (300 eV case at $z=13$, Fig
\ref{fig:power-spectrum}).
If we imagine a full spectrum of those peaks (of different mass for
example), it is possible that the Ly$\alpha$ pumping field has nonlinear
fluctuations around the mean at $\left\langle x_{\alpha}\right\rangle \sim1$.
Many models already find that this is indeed the case, where
astrophysical information dominates the fluctuating signal over the
cosmological information even during the CD.
Second, the UV emissivity of minihalos inside our simulation is
relatively small mainly due to the effective minimum mass of
minihalos hosting stars is large, which is $\sim 3\times 10^{6}\,M_\odot$.
The UV feedback may be much more advanced
than what has been simulated here, as is inferred by e.g. \citet{Ahn2012}
where the mean ionization fraction at $z\sim15$ can be as high as
$\sim24\%$ if efficient formation of Pop III stars in minihalos
of $M\ge10^{5}\, M_{\odot}$ is assumed, even with Lyman-Werner feedback
on the primordial cooling agent ${\rm H}_{2}$. 
The 21-cm signal will
then be again dominated by the astrophysical information, 
whose observation would be more useful in astrophysics, such as
discriminating between these two models with very different 
efficiencies in Pop-III star formation.
Thus for 21-cm
cosmology, it is preferred that the observation is done at the highest
possible redshift to enable at least the $\mu$-decomposition of the
observed power spectrum (e.g. \citealt{2005ApJ...624L..65B}; MSMIKA)
when every field relevant for building up the 21-cm $P(k)$ is in the
linear regime. The baryon-dark matter offset velocity (\citealt{Tseliakhovich2010})
is in some sense another nuisance to cosmology even at very high redshift,
because the rate and spatial fluctuations of the induced shock heating
are not yet well understood, even though the power spectrum can be strongly boosted
for much easier observation than what had been predicted before (\citealt{McQuinn2012}).

\acknowledgments
We thank Andrei Mesinger and Xuelei Chen for helpful
  comments. This work was supported by a research grant from Chosun University (2009).

\appendix
\section*{transfer of 21 cm radiation with finite optical depth
and frequency overlap}
In general, the nonlinear evolution of density and temperature 
will generate non-negligible opacity to the 21-cm radiation from place
to place. In
addition, the peculiar velocity of gas elements shifts
the observing frequency and can
make photons go through the 21-cm transition multiple times along the
LoS while ending up with
the same observing frequency. Therefore, even though the line optical depth
is defined locally as an integrated quantity of a Dirac delta function
under cosmological scale, we need to carefully treat the
multi-valuedness of line transitions as follows. It becomes the most
convenient when one considers uniformly discretized redshift 
space, and map the real-space mesh into the redshift-space mesh, which
was dubbed as mesh-to-mesh (MM) real-to-redshift-space-mapping (RRM) scheme
in MSMIKA. While MSMIKA show 
explicitly how MM-RRM under frequency overlap can be done in the optically thin limit, it was
not shown explicitly how to perform MM-RRM in the generic case with
finite optical depth and frequency overlap. Here we describe the detailed
scheme for the MM-RRM and the radiative transfer in the most
generic cases.

The most generic form of the
transfer equation (Equation 18 of MSMIKA) is
\begin{equation}
I_{\nu_{\rm obs}}=I_{\nu_{\rm obs}}^{\rm CMB} 
{\rm e}^{-\tau_{\nu_{\rm obs}}} +\int_{0}^{\tau_{\nu_{\rm obs}}}S_{\nu'}(\xi')
{\rm e}^{-(\tau_{\nu_{\rm obs}}-\tau'_{\nu'})}d\tau'_{\nu'},
\label{eq:genericInu}
\end{equation}
which can also be expressed in terms of brightness temperatures:
\begin{equation}
T_{b}(\nu_{\rm obs})=T_{\rm CMB,\,0} 
{\rm e}^{-\tau_{\nu_{\rm obs}}} +\int_{0}^{\tau_{\nu_{\rm obs}}}
\frac{T'_{\rm S}}{1+z'}\left(1-\frac{v'_{\parallel}}{c}\right)
{\rm e}^{-(\tau_{\nu_{\rm obs}}-\tau'_{\nu'})}d\tau'_{\nu'},
\label{eq:genericT}
\end{equation}
when Equations (28) and (29) of MSMIKA are used. Here $z'$ is the
cosmological redshift of a gas element, which is shifted in the
redshift space (or the observing frequency space) due to the LoS
peculiar velocity $v'_{\parallel}$, with $\nu_{\rm obs}=\nu'
(1-v'_{\parallel}/c)/(1+z')$. The variance in $v'_{\parallel}$ thus can
make different gas elements along the LoS to be observed at the same
$\nu_{\rm obs}$ (frequency overlap). Due to  thermal broadening,
this overlap may be 
extended a little bit across more gas elements, but we find that the
peculiar velocity is the dominant cause of the frequency overlap,
and thus ignore  thermal
broadening. As shown in MSMIKA, correction to the optical depth
due to thermal broadening at about the highest temperature
reached, $T=10^4 \rm K$, is only of order $\sim 10^{-9}$.

We can write Equation~(\ref{eq:genericT}) more explicitly. If there are
N locations along the LoS yielding the same $\nu_{\rm obs}$,
Equation~(\ref{eq:genericT}) becomes
\begin{equation}
T_{b}(\nu_{\rm obs})=T_{\rm CMB,\,0} 
{\rm e}^{-\tau_{\nu_{\rm obs}}} + \sum_{i=1}^{N}
\frac{T_{{\rm S},i}}{1+z_{i}}\left(1-\frac{v_{\parallel,i}}{c}\right)
({\rm e}^{\tau_{i}}-1)
{\rm e}^{-(\tau_{\nu_{\rm obs}}-\sum_{j=1}^{i-1}\tau_{j})},
\label{eq:genericTsolved}
\end{equation}
where indicies $i$ and $j$ increase toward the near side of the
simulation box to the
observer, a gas element at $z_i$ has the optical depth
\begin{equation}
\tau_{i}=\frac{3c^3 A_{10}T_{*}n_{\rm HI}(z_i)}{32\pi
  \nu_{0}^{3}T_{\rm S}(z_i)|H(z_{i})/(1+z_{i})+dv_{\parallel,i}/dr_{\parallel}|(1-v_{\parallel,i}/c)},
\label{eq:tau_i}
\end{equation}
and the total optical depth of all the gas elements with the same
$\nu_{\rm obs}$ is $\tau_{\nu_{\rm obs}}\equiv \sum_{i=1}^{N}\tau_i$. A gas
element at $z_i$ emits radiation in proportion to 
(${\rm e}^{\tau_{i}}-1)$, and is attenuated by the accumulated optical depth
$\tau_{\nu_{\rm obs}}-\sum_{j=1}^{i-1}\tau_{j}=\sum_{j=i}^{N}\tau_{j}$
by those ``in front of'' the
element and the element itself. The differential brightness
temperature at
$\nu_{\rm obs}$ then becomes
\begin{equation}
\delta T_{b}(\nu_{\rm obs})= 
\sum_{i=1}^{N}
\frac{T_{{\rm S},i}}{1+z_{i}}\left(1-\frac{v_{\parallel,i}}{c}\right)
(1-{\rm e}^{-\tau_{i}})
{\rm e}^{-\sum_{j=i+1}^{N}\tau_{j}}
- T_{\rm CMB,\,0} (1 - {\rm e}^{-\tau_{\nu_{\rm obs}}}),
\label{eq:dtb_gen}
\end{equation}
which obviously converges to the usual form (e.g. Equation 47 of MSMIKA)
for the case without frequency overlap, when $N=1$.

For MM-RRM, we first form an empty data cube
corresponding to the simulation box at the
redshift of interest but with the LoS observing frequency as the LoS
axis. Without peculiar velocities, then, there will be a one-to-one
mapping of the real-space cube onto the observing-space cube. In
general, however, we need to shift each cell according to the peculiar
velocity. We shift each LoS boundary between two adjacent real-space cells 
in the redshift space by the average of the two
cell-centered peculiar velocities in addition to the cosmological
redshift. This guarantees that these cells do not 
overlap in frequency unless a frequency-crossing of two boundaries
occurs. The same frequency shifting scheme is used in e.g.
\citep{Mellema2006a} and MSMIKA.

\begin{figure}
\begin{center}

\includegraphics[width=0.7\textwidth]{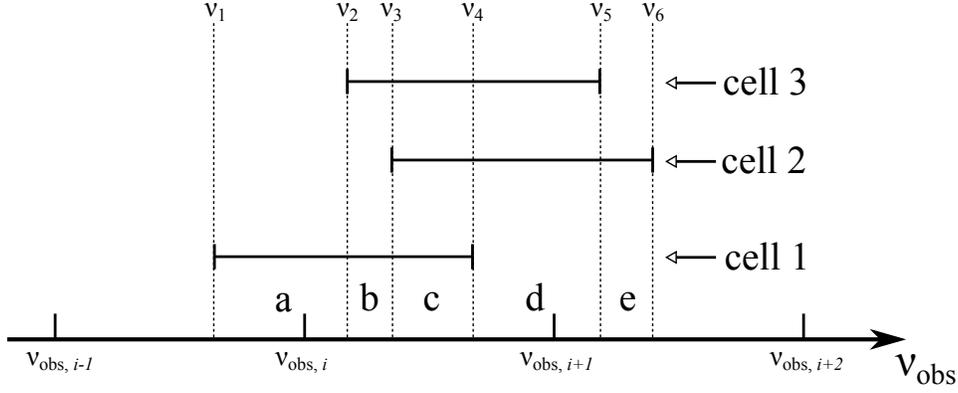}
\end{center}

\caption{ Illustration of how we transfer the 21-cm line
  radiation and map the signal to the redshift (or observing
  frequency) space. Real-space cells are shifted in the observing
  frequency  ($\nu_{\rm obs}$)
  space, and three cells (cells 1, 2 and 3) are mutually overlapping around
  $\nu_{\rm obs}=[\nu_{{\rm obs},i},\,\nu_{{\rm obs},i+1}]$. Total of
  5 ``bands'' (a, b, c, d, and e) form, and they bear 
  cumulative optical depths different from one another. $\delta T_{b}$
  is calculated for each band using Equation~(\ref{eq:dtb_gen}), and is
  weighted by the ratio of the width of the band to the width of the
  observing bin.
\label{fig:method}}
\end{figure}

As depicted in Figure~\ref{fig:method}, if there are $N$
frequency-overlapping cells along a LoS, there arise $2N-1$
(observing) frequency bands ($\nu_{\rm obs}=[\nu_{{\rm obs},1}, \,\nu_{{\rm obs},2}]$,
$[\nu_{{\rm obs},2}, \,\nu_{{\rm obs},3}]$, $\cdots$, 
$[\nu_{{\rm obs},2N-1}, \,\nu_{{\rm obs},2N}]$) with mutually different cumulative
optical depths. For example, the 5 bands (a, b, c, d, and e) in
Figure~\ref{fig:method} have cumulative optical depths $\tau_1$, $\tau_1
+\tau_3$, $\tau_1 +\tau_2 +\tau_3$, $\tau_2 + \tau_3$, and $\tau_2$,
respectively. The number of 21cm transitions and the net contribution
to a relevant $\nu_{\rm obs}$-bin vary over these bands as well:
e.g. inside the band b, 
\begin{eqnarray}
\delta T_{b}({\rm band}\,\,b)= 
\frac{T_{{\rm S},1}}{1+z_{1}}\left(1-\frac{v_{\parallel,1}}{c}\right)
(1-{\rm e}^{-\tau_{1}})\,{\rm e}^{-\tau_{3}} 
+\frac{T_{{\rm S},3}}{1+z_{3}}\left(1-\frac{v_{\parallel,3}}{c}\right)
(1-{\rm e}^{-\tau_{3}})
- T_{\rm CMB,\,0} (1 - {\rm e}^{-\tau_{1}-\tau_{3}}),
\label{eq:dtb_exam}
\end{eqnarray}
and its contribution to the observing bin of 
$\nu_{\rm obs}=[\nu_{{\rm obs},i},\,\nu_{{\rm obs},i+1}]$ becomes  
$f({\rm band \,\,b})\,\delta T_{b}({\rm band \,\,b})$,
where $f({\rm band \,\,b})\equiv(\nu_{3}-\nu_{2})/(\nu_{{\rm
    obs},i+1}-\nu_{{\rm obs},i})$. Inside each band, only one
radiation transfer calculation is required, negating the need for further
multi-band calculation.
Of course, bands that extend across two or
more observing bins (e.g. band a and d) should be further split into
the ones belonging to
the individual bins. 

\begin{figure}
\begin{center}

\includegraphics[width=0.4\textwidth]{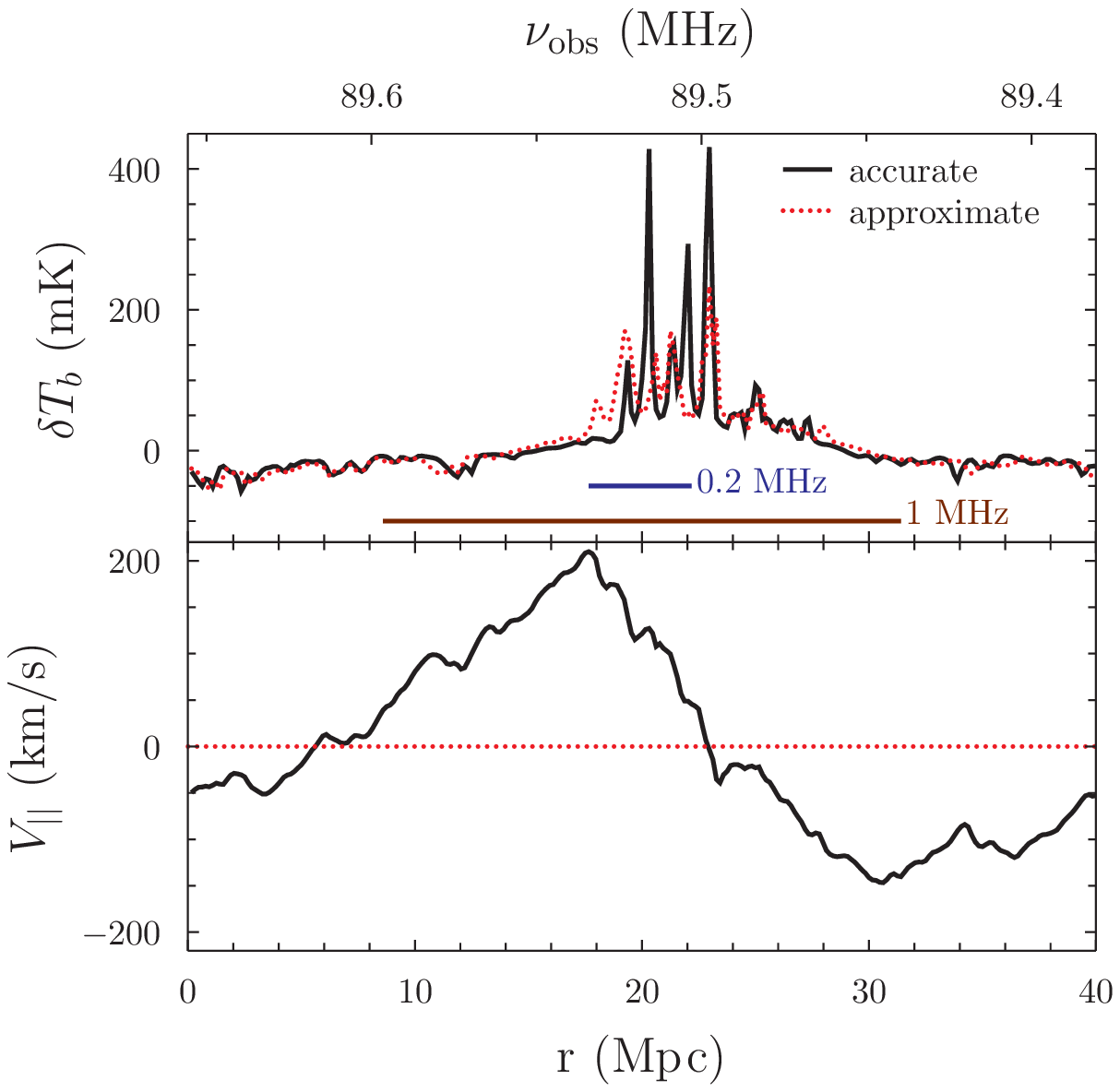}
\end{center}

\caption{ (top) Comparison of accurately calculated $\delta T_{b}$
  (black, solid) and approximated $\delta T_{b}$ (red, dotted), from a
  box at $z=15$, for the model with the power-law X-ray SED 
  along a LoS same as the ones in Figure~\ref{fig:densvel}. The 
  accurate calculation considers both the finite optical depth and the
  effect of the peculiar velocity, while the approximate calculation
  uses the optically thin approximation and ignores the peculiar
  velocity.  They are plotted against both the (comoving)
  length scale of the box ($r$) and the observed frequency ($\nu_{\rm
    obs}$). Also shown are the bandwidths of the frequency-integration
  we consider:
  0.2 MHz (blue, solid) and 1 MHz (brown, solid). (bottom) Peculiar
  velocity of gas elements projected along the LoS. 
\label{fig:dtbcompare}}
\end{figure}

We use cell-centered or
cell-averaged values for all the relevant quantities in 
Equations (\ref{eq:tau_i}) and (\ref{eq:dtb_gen}) and calculate 
$\delta T_{b} ({\rm band\,\,k})$.
Especially, we let
$dv_{\parallel,i}/dr_{\parallel}=(v_{\parallel,i-1}-v_{\parallel,i+1})/(2\Delta
r_{\parallel, {\rm cell}})$, in accordance with how we shift cell
boundaries. 
The average $\delta T_{b}$ inside each observing bin can
be calculated as
$\delta T_{b}=\sum_{k}f({\rm  band\,\,k})\,\delta T_{b} ({\rm
  band\,\,k})$. Iterating this process over all the bands, one can
fill the full data cube and finish 
MM-RRM with a full radiative transfer.

We find that the optically thin approximation works reasonably well,
even in this highly nonlinear problem, 
because most cells indeed have small optical depths. With the maximum
$v_{\parallel}$ reaching $\sim 200$ km/s, however, there exists
non-negligible shift of $\delta T_{b}$ in the redshift space, which
will yield the mixture of the locally increased (when
frequency-overlap occurs) and
decreased (when some frequency-gap occurs) signals
(Figure~\ref{fig:dtbcompare}). Nevertheless, with the frequency filter $\Delta \nu
\gtrsim 0.2\,{\rm MHz}$, the frequency-integrated signal is not too strongly
affected in our case. We find that in all the cases we examined, a
maximum of $\sim 10\%$ difference occurs between the accurate
calculation and the approximate one.

%%The practical algorithm is as follows:
%%\begin{enumerate}
%%\item Let $k$ be the index of real-space cells along a LoS, running
%%  from 1 at the far side from the observer to $N_{\rm cell, LoS}$ (=256 in our case)
%%  at the near side to the observer.
%%\item Shift each LoS boundary between two adjacent real-space cells 
%%in the redshift space by the average of the two
%%cell-centered peculir velocities in addition to the cosmological
%%redshift. This guarantees that these cells do not 
%%overlap in frequency unless a shell-crossing
%%occurs. 
%%\item Start from $k=1$ and see how many cells in front (toward the
%%  observer) overlap with the cell in frequency space. If $N-1$ cells
%%  overlap with the cell, mark the observing frequecies made by cell
%%  boundaries which would naturally form $2N-1$ frequency bands as
%%  depicted in Figure~\ref{fig:method}.
%%\item For each band, 
%%\end{enumerate}

%\bibliographystyle{apj}
%\bibliography{refs_2014dec}

\end{document}